\documentclass[structabstract]{aa} 
 \usepackage{longtable}
\usepackage{epsfig}
\usepackage{graphicx}
\usepackage{txfonts}
\usepackage{amssymb}
\usepackage{natbib}                          
\bibpunct{(}{)}{;}{a}{}{,}

\newcommand{\be}{\begin{equation}}
\newcommand{\ee}{\end{equation}}
\newcommand{\bea}{\begin{eqnarray}}
\newcommand{\eea}{\end{eqnarray}}
\newcommand{\fc}{f_{\rm c}}

\newcommand{\leddth}{L_{\rm Edd}}
\newcommand{\feddth}{F_{\rm Edd}}
\newcommand{\feddst}{F_{\rm Edd}^*}
\newcommand{\teddth}{T_{\rm Edd}}
\newcommand{\teddinf}{T_{\rm Edd,\infty}}

\newcommand{\kappae}{\kappa_{\rm e}}
\newcommand{\kappaR}{\kappa_{\rm R}}

\newcommand{\rmd}{{d}}
\newcommand{\fourx}{\underline{x}}
\newcommand{\fourp}{\underline{p}}
\newcommand{\vecx}{\mbox{\boldmath $x$}}
\newcommand{\vecp}{\mbox{\boldmath $p$}}
\newcommand{\vn}{\mbox{\boldmath $n$}}
\newcommand{\vomega}{\mbox{\boldmath $\omega$}}
\newcommand{\vOmega}{\mbox{\boldmath $\Omega$}}

\newcommand{\unb}{\underline{\nabla}}
\newcommand{\vnabla}{\mbox{\boldmath $\nabla$}}

\newcommand{\noccx}{n} 
\newcommand{\nocce}{\tilde{n}_{\rm e}}
\newcommand{\re}{r_{\rm e}}
\newcommand{\fe}{f_{\rm e}}
\newcommand{\me}{m_{\rm e}}
\newcommand{\Ne}{N_{\rm e}}
\newcommand{\Te}{T_{\rm e}}
\newcommand{\lambdac}{\lambda_{\rm C}}
\newcommand{\sigmat}{\sigma_{\rm T}}
\newcommand{\taut}{\tau_{\rm T}}

\begin{document}

 \title{X-ray bursting neutron star  atmosphere models 
 using an exact relativistic kinetic equation for Compton scattering}

\author{
V.~Suleimanov\inst{1,2} 
\and
J.~Poutanen\inst{3} 
\and
K.~Werner\inst{1}}

 
\institute{
Institut f\"ur Astronomie und Astrophysik, Kepler Center for Astro and
Particle Physics,
Universit\"at T\"ubingen, Sand 1,
 72076 T\"ubingen, Germany \\
   \email{suleimanov@astro.uni-tuebingen.de,werner@astro.uni-tuebingen.de}
\and
Kazan Federal University, Kremlevskaja str., 18, Kazan 420008, Russia
         \and
Astronomy Division, Department of Physics, P.O. Box 3000, FI-90014 University of Oulu, Finland \\
       \email{juri.poutanen@oulu.fi}
}

\date{Received 25 April 2012 / Accepted 13 July 2012}

   \authorrunning{Suleimanov et al.}
   \titlerunning{X-ray bursting neutron star  atmosphere models}

\abstract
{Theoretical spectra of X-ray bursting neutron star (NS) model atmospheres are widely used to 
determine the basic NS parameters such as their masses and radii. 
Compton scattering, which plays an important role in spectra formation at  high luminosities, 
is often accounted for using  the differential Kompaneets operator, 
while in other models a more general,  integral operator for the Compton scattering kernel is used. }
{We  construct  accurate NS atmosphere models using for the first time 
an exact treatment of Compton scattering via the integral relativistic  kinetic equation.
We also  test various approximations to the Compton scattering redistribution function and 
compare the results with the previous calculations based on the Kompaneets operator.  }
{We solve the radiation transfer equation together with the hydrostatic equilibrium equation accounting 
exactly for the radiation pressure by electron scattering. 
We use the exact relativistic  angle-dependent redistribution function as well as its simple approximate representations.  
}
{We thus construct a new set of plane-parallel atmosphere models in local thermodynamic equilibrium
 (LTE) for hot NSs. 
The models were computed for six chemical compositions (pure H, pure He, solar H/He mix with various heavy elements abundances 
$Z$ = 1, 0.3, 0.1, and 0.01$Z_{\odot}$, and three surface gravities $\log g$ = 14.0, 14.3, and 14.6.  
For each chemical composition and surface gravity, 
we compute more than 26 model atmospheres with various  luminosities relative to the  
Eddington luminosity $\leddth$ computed for the Thomson cross-section. 
The maximum relative luminosities $L/\leddth$ reach values of up to 1.1 for  high gravity models.
The emergent spectra of all models are redshifted and fitted by diluted blackbody spectra in the 3--20 keV energy range 
appropriate for the {\it RXTE}/PCA. 
We also compute the color correction factors $\fc$.
}
{The radiative acceleration $g_{\rm rad}$ in our luminous, hot-atmosphere models is significantly smaller than 
in corresponding models based on the Kompaneets operator, because of the Klein-Nishina reduction 
of the electron scattering cross-section, 
and therefore formally ``super-Eddington'' model atmospheres do exist. 
The differences between the new and old model atmospheres are small for $L/\leddth< 0.8$.  
For the same $g_{\rm rad}/g$, the new $\fc$ are slightly larger (by approximately 1\%)  than the old values. 
We also find that the model atmospheres,  the emergent spectra, and the color correction factor 
computed using angle-averaged and approximate Compton scattering kernels differ from the exact solutions by less than 2\%. 
}


\maketitle
%

\section{Introduction}

X-ray bursting neutron stars (NSs) are members of low-mass X-ray binaries with 
quasi-periodical thermonuclear flashes on 
their surfaces \citep[see reviews by][]{LvPT93, SB06}.
Thermonuclear burning occurs  at the bottom of the freshly accreted matter and can be 
so powerful that the luminosity 
reaches the Eddington limit. 
These bursts lead to photospheric radius expansion (PRE) and are a potentially 
powerful tool for determining  the NS masses and radii \citep{Ebisuzaki87,Damen:90,vP:90}. 
The knowledge of NS basic parameters is extremely important for establishing the  physical properties
 (equation of state) of supra-nuclear  dense matter in the NS inner cores  \citep[see][for a review]{LP07}.

The precise radius determination of the NSs from X-ray bursts is  impossible without accurate spectral models. 
The observed spectra of X-ray bursts are often well-fitted by a blackbody  \citep{gallow08}. 
Theoretical models of hot NS atmospheres show  \citep{Londonetal:86,Lapidus:86,Pavlov.etal:91} 
that the emergent spectra are close to diluted blackbody spectra 
$F_{E} \approx \fc^{-4} B_{E} (\fc T_{\rm eff})$ 
owing to strong energy exchange caused by Compton scattering. 
The color correction factor $\fc \equiv T_{\rm c}/ T_{\rm eff}$,
which relates the color temperature of the spectrum $T_{\rm c}$ to the  effective temperature of the 
atmosphere $T_{\rm eff}$, takes values of about 1.3--1.9. 
Theoretical dependences of the color correction factor on luminosity 
for various chemical compositions and gravities were computed by  \citet[][hereafter SPW11]{SPW11}.
The atmosphere models used in their work were based  on the Kompaneets operator  
\citep{Kompaneets:57}  description of Compton scattering. 
This approach was also used in most previous studies \citep{Londonetal:86,Lapidus:86,Ebisuzaki87,Pavlov.etal:91}.
The Kompaneets operator describes Compton scattering in the 
non-relativistic, isotropic, diffusion approximation, 
which should still be rather adequate for 
hot NS model atmospheres with typical effective temperatures below $\sim$2 keV. 
On the other hand, Madej and collaborators \citep{madej:91,Madej.etal:04,Majczyna:05}  
used the integral description of Compton scattering employing the angle-averaged redistribution function (RF) 
derived by \citet{Guilbert81}.
A fully relativistic treatment of Compton scattering becomes important at luminosities 
close to the Eddington values, because of the reduction of the 
effective cross-section and the corresponding drop in the radiation pressure force. 
This obviously changes the maximum value of the luminosity where hydrostatic 
equilibrium can still be achieved, i.e. the actual 
value of the Eddington luminosity. 
This has important implications for the method of determining the NS masses and radii from PRE bursts,  which 
is based on the Eddington luminosity (see \citealt{S11}, hereafter S11). 
It is therefore,  necessary to check the validity of model atmospheres and spectral properties 
computed with the Kompaneets operator by comparing the results with more accurate models 
based on an exact treatment of Compton scattering  using relativistic kinetic equations. 
 
Here we construct new models of hot NS atmospheres based on an exact treatment of Compton scattering 
that employs a fully relativistic, angle-dependent RF  \citep{AA81,PKB86,NP94,PS96}.  
In Sect. \ref{s:methods}, we  present the main equations describing the atmosphere models as well as
the methods for their solution. 
We also compare  atmosphere models based on various RFs for Compton scattering and 
using the Kompaneets operator.  
In Sect. \ref{s:newgrid}, we present a new set of atmosphere models, 
emergent spectra, and the color correction factors. 
We compare them to previous results obtained with the Kompaneets operator.  
In Sect.~\ref{s:application}, we discuss applications of the new models to the observations of X-ray bursts  
and determination of the NS masses and radii. 
We summarize in Sect.~\ref{sec:summary}. 
In Appendix~\ref{sec:rke} we derive the exact RF for Compton scattering. 
The details of the method of solution of the radiative transfer equation are presented in Appendix~\ref{sec:method_rte}. 
Comparison with the previous attempts to compute atmosphere models using 
integral approach to Compton scattering are given in Appendix~\ref{sec:madej}. 
And, finally, Appendix~\ref{app:D} presents the spectral characteristics of the models. 

\section{Method of atmosphere modeling}
\label{s:methods}  

\subsection{Main equations} 

A  method for modeling hot X-ray bursting NS atmospheres that employs the Kompaneets operator for 
Compton scattering  was  described in detail in SPW11. 
Here we repeat the basic assumptions and emphasize the differences arising because of
the new treatment of Compton scattering as well as radiation transfer. 

We compute models in hydrostatic and radiative equilibrium in a plane-parallel approximation. 
The main input parameters are the chemical composition (particularly the hydrogen mass fraction $X$), 
the surface gravity 
\be \label{eq:g_def}
   g=\frac{GM}{R^2}(1+z),
\ee
and the relative NS luminosity $l = L/\leddth$,
where  $\leddth$ is the Eddington luminosity  measured at the NS surface
\be \label{eq:ledd_def}
   \leddth=\frac{4\pi GMc}{\kappae} (1+z), 
\ee
for the Thomson scattering opacity 
\be \label{eq:kappae}
\kappae = \sigmat \frac{\Ne}{\rho} \approx 0.2\ (1+X) \ \mbox{cm}^2\ \mbox{g}^{-1} . 
\ee
Here $\sigmat= 6.65\times 10^{-25}$ cm$^2$ is the Thomson cross-section, 
$\rho$ is the gas density, and $\Ne$  is the electron number density.
The gravitational redshift is related to the NS parameters as
\be \label{eq:redshift_def}
    1+z=(1-2GM/c^2R)^{-1/2} .
\ee
We assume that the flux is constant throughout the NS surface. 
Our calculations are valid for a patch on the NS surface if instead of the relative luminosity 
we consider the relative flux, $l = F/\feddst$, as a parameter, where  
\be\label{eq:fedd_def}
\feddst= \frac{\leddth}{4\pi R^2} = \frac{GMc}{R^2 \kappae} (1+z) . 
\ee
The effective temperature $T_{\rm eff}$ can be expressed via $l$ as 
\be \label{eq:teff}
   T_{\rm eff} = l^{1/4} \teddth,
\ee
where the Eddington temperature $\teddth$ is the maximum possible effective temperature on the NS surface, 
which is evaluated using the Thomson scattering opacity
\be \label {eq:tedd}
   \sigma_{\rm SB} \teddth^4 = \feddst = \frac{gc}{\kappae} .
\ee

The structure of the atmosphere  for an X-ray bursting NS 
is described by a set of differential equations. The first one is the hydrostatic equilibrium equation  
\be \label{eq:hyd}
  \frac {d P_{\rm g}}{dm} = g - g_{\rm rad},
\ee
where $g_{\rm rad}$ is the radiative acceleration 
and $P_{\rm g}$  is the gas pressure and the column density $m$ is defined as
\be
\rmd m = -\rho \, \rmd s \, ,
\ee
where $s$ is the vertical distance.

The second equation is the radiation transfer equation for the specific intensity $I(x,\mu)$
accounting for Compton scattering (see Appendix \ref{sec:rke} for derivation). 
In the plane-parallel approximation, it has the form
\be \label{eq:rte}
\mu \frac{\rmd I(x,\mu)}{\rmd\tau (x,\mu)} = I(x,\mu) - S(x,\mu),
\ee
where 
\be
    \rmd\tau (x,\mu) = \left[\sigma(x,\mu)+k(x)\right]\, \rmd m ,
\ee
$\mu = \cos \theta$ is the cosine of the angle between the surface normal  and the direction 
of radiation propagation, 
$x=h\nu/m_{\rm e}c^2$ is the photon energy in units of electron rest mass,  and $k(x)$ is  the
``true'' absorption opacity. 
 The electron scattering opacity accounting for the  induced scattering is
\be
\label{eq:scatopac}
  \sigma(x,\mu) \! = \!  \kappae \frac{1}{x} \!\int\limits_0^\infty\! x_1 \rmd x_1
 \!\!\!  \int\limits_{-1}^1 \!\!\rmd\mu_1 R(x_1,\mu_1;x,\mu)\,  
\! \left(1\!+\!\frac{C\,I(x_1,\mu_1)}{x_1^3}\right),
\ee
and 
\be \label{eq:const_int}
C= \frac{1}{2m_{\rm e}} \left( \frac{h}{\me c^2} \right)^3 . 
\ee
The source function is a sum of the thermal part and the scattering part
\bea \label{eq:source}
\lefteqn{ S(x,\mu)  =   \frac{k(x)}{\sigma(x,\mu)+k(x)} \, B_x +
\frac{\kappae}{\sigma(x,\mu)+k(x)}   
}  \\  
&\times &   \left(1+\frac{C\,I(x,\mu)}{x^3}\right)  x^2 \int_0^\infty
\frac{dx_{\rm 1}}{x^2_{\rm 1}} \int_{-1}^1 d\mu_{\rm 1} R(x,\mu;x_{\rm 1}, \mu_1) I(x_{\rm 1}, 
\mu_{\rm 1}), \nonumber
\eea
where 
\be \label{eq:planck}
B_x = B_{\nu} \frac{\rmd \nu}{\rmd x}   = \frac{x^3}{C} \frac{1}{\exp(x/\Theta)-1}, 
\ee  
with $B_{\nu}$ being the Planck function and 
\be \label{ipc_u8}
 \Theta = \frac{kT}{\me c^2}
\ee
the electron temperature. 

The RF $R(x,\mu;x_1,\mu_1)$ describes the probability that a photon with the dimensionless energy $x_1$ 
propagating in the direction  corresponding to $\mu_1$ is scattered to energy $x$ and in a
 direction corresponding to $\mu$.  
This function is found by integrating over the azimuthal angle $\varphi$ of the RF $R(x,x_1,\eta$), 
which depends on the cosine of the angle between the 
directions of the photon propagation before and after scattering $\eta$
\bea \label{eq:rxxmu}
R(x,\mu;x_1,\mu_1) = \int_0^{2\pi} R(x,x_1,\eta)\,d\varphi,\\ \nonumber 
\eta=\mu\mu_1+\sqrt{1-\mu^2}\sqrt{1-\mu_1^2}\cos\varphi.
\eea
The  RF depends on the depth $s$ through the electron temperature 
and satisfies the relation (see Eq. \ref{eq:rf_symm}) 
\be
R(x_1,\mu_1;x,\mu) =  R(x,\mu;x_1,\mu_1)\, \exp\left(\frac{x-x_1}{\Theta}\right) ,
\ee
which is the consequence of the detailed balance relation \citep[][see Appendix \ref{sec:rke}]{Pom73,NP94}.
This  implies that the source function given by Eq. (\ref{eq:source}) equals the Planck function 
for the photon field described by the Bose-Einstein distribution of any chemical potential. 

In this paper, we use two  RFs for Compton scattering. 
The first one is the fully relativistic exact RF valid for any photon energy and electron temperature
\citep[][see Eqs. \ref{eq:red_phi} and \ref{eq:r0f} in Appendix \ref{sec:rf}]{AA81,NP93,NP94,PS96}. 
The second one is an approximate RF corresponding to the isotropic scattering in the electron rest frame 
(Eq. \ref{eq:r0tot_app} in Appendix \ref{sec:rf}),  
which is accurate at temperatures below about 100 keV and non-relativistic photon energies  
\citep{an80,Pou94PhD,PS96}. 
In both cases, we also consider the angle-averaged RFs  
\be  \label{eq:rxx_ave}
   R(x,x_1) = \frac{1}{2} \int_{-1}^{+1}\,\rmd\eta \,R(x,x_1,\eta) ,
\ee
substituting it instead of the angle-dependent RF into Eq. (\ref{eq:rxxmu}). 
 
The formal solution of the radiation transfer equation (\ref{eq:rte})
is obtained using the short-characteristic method  \citep{OK87} in three angles in each hemisphere,
and the full solution is found with an accelerated $\Lambda$-iteration method (see details in Appendix 
\ref{sec:method_rte}).

The radiation pressure acceleration $g_{\rm rad}$ is computed using the RF as 
\bea \label{eq:grad}
&&g_{\rm rad} = \frac{\rmd P_{\rm rad}}{\rmd m} = 
\frac{2\pi}{c} \,\frac{\rmd}{\rmd m}\, \int^{\infty}_{0} \rmd x \, \int^{+1}_{-1}\mu^2 I(x,\mu)\, \rmd\mu \\ 
\nonumber
&&= \frac{2\pi}{c} \int^{\infty}_{0} \!\! \rmd x  \int^{+1}_{-1} \left[\sigma(x,\mu) + k(x)\right]  
\left[I(x,\mu)-S(x,\mu)\right] \,\, \mu \, \rmd\mu, 
\eea
where the derivative with respect to $m$ is replaced by the first moment of the
 radiation transfer equation (\ref{eq:rte}). 
When the  source functions and the opacities are isotropic, this expression is reduced to the standard definition 
\be \label{eq:grad_stan}
g_{\rm rad} = \frac{4\pi}{c} \, 
\int^{\infty}_{0}   \left[\sigma(x) + k(x)\right]   H_x(m) \ \rmd x , 
\ee
where 
\be
     H_x =\frac{1}{2}\, \int^{+1}_{-1}\mu \, I(x,\mu)\, \rmd\mu 
\ee
is the first moment of specific intensity. 
 
These equations are completed  by the energy balance equation
\be  \label{eq:econs}
\int^{\infty}_{0} dx \, \int^{+1}_{-1} \left[\sigma(x,\mu) + k(x)\right]  \left[I(x,\mu)-S(x,\mu)\right] \, \, d\mu = 0,
\ee
the ideal gas law
\be   \label{gstat}
    P_{\rm g} = N_{\rm tot}\ kT,
\ee
where $N_{\rm tot}$ is the number density of all particles, and  the particle and charge conservation equations.  
In our calculations, we assumed local thermodynamic equilibrium (LTE), therefore 
the number densities of all ionization and excitation states of all elements were
calculated using the Boltzmann and Saha equations. 
We accounted for  the pressure ionization effects  on hydrogen and helium populations using the occupation
probability formalism \citep{Hum.Mih:88} as  described by \citet{Lanz.Hub:94}. 
In addition to electron scattering,  we took into account the free-free opacity 
as well as the bound-free transitions  
for all ions of the 15 most abundant chemical elements (H, He, C, N, O, Ne, Na, Mg, Al, Si, S, Ar, Ca, Fe, Ni) \citep[see][]{Ibragimov.etal:03} 
using opacities from \cite{VYa:95}.

\subsection{Method of solution} 

To solve the above equations, we
used our version of the computer code ATLAS \citep{Kurucz:70,Kurucz:93},
modified to deal with high temperatures \citep{Sul.Pout:06,sw:07}.   The code was 
further developed to account for Compton scattering using the RF approach.

In our computations, we used 300--360 logarithmically equidistant frequency points in the range 
$10^{14}$--$10^{20}$ Hz  ($\approx 4 \times 10^{-4}$--400 keV) for the 
luminous model atmospheres ($l  \geq 0.1$), 
and $10^{14}$--$10^{19}$ Hz for $l  < 0.1$.
The calculations were performed at a set of 98 depth points $m_{\rm i}$ 
distributed equidistantly on the logarithmic scale from $10^{-6}$ to $m_{\rm max}=10^6$~g~cm$^{-2}$. 
The appropriate value of $m_{\rm max}$ was chosen to satisfy the condition $\sqrt{\tau_{\nu,\rm
b-f,f-f}(m_{\rm max})\tau_{\nu}(m_{\rm max})} >$ 1 at all frequencies,
where $\tau_{\nu,\rm b-f,f-f}$ is the optical depth computed with the true
opacity  only (bound-free and free-free transitions, without scattering).  
This requirement was necessary for satisfying the inner boundary condition
of the radiation transfer problem.

\begin{figure} 
\begin{center}
\includegraphics[width= 0.73\columnwidth]{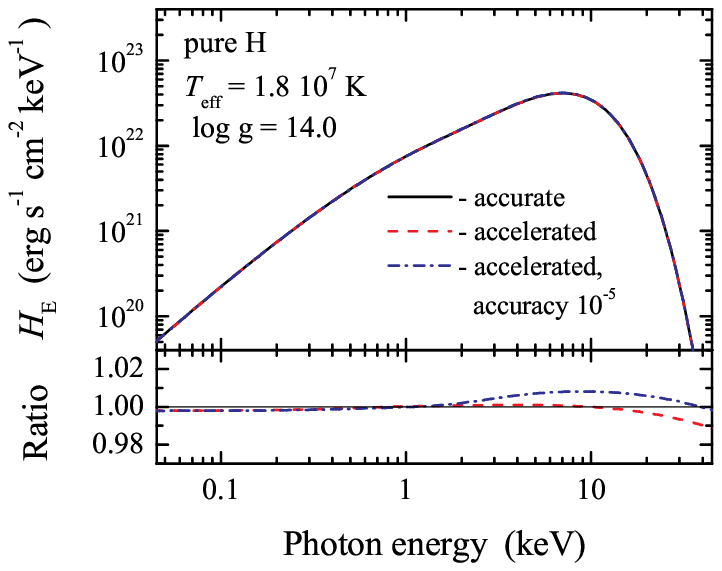}
\includegraphics[width= 0.73\columnwidth]{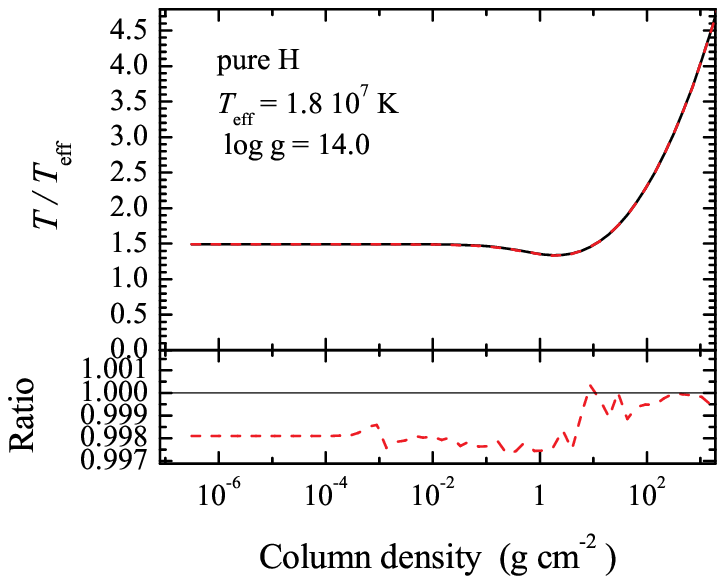}
\caption{\label{fig:1}
Emergent spectrum ({\it top panel}) and temperature structure ({\it bottom panel}) of the fiducial model  
(pure hydrogen,  $T_{\rm eff} = 1.8\times10^7$ K, $\log g$ = 14.0)  computed using three different methods. 
In accurate method 1 every $\Lambda$-iteration starts from the thermal part of the source function
(solid curves give the results for  the relative accuracy of $10^{-4}$).  
In accelerated  method 2, the $\Lambda$-iterations start from the source function taken 
from previous temperature iteration, but at every fifth temperature correction they start from the thermal part 
of the source function (dashed curves show results for  the relative accuracy of $10^{-4}$). 
Method 3 is the same as method 2, but for the relative accuracy of $10^{-5}$ (dot-dashed curve). 
In the {\it top lower panel}, the ratios of the spectra for methods 2 and 3
to the spectrum computed with method 1 are shown. 
The ratio of  the temperature structures computed using methods 2 and 1 
is shown in the {\it bottom lower panel}. 
} 
\end{center} 
\end{figure}

The course of the calculations was the same as for the method that adopts the Kompaneets operator (SPW11).  
First, a starting grey atmosphere model was calculated and opacities at all
depth points and all frequencies  were obtained. 
The solution of the radiative transfer equation (\ref{eq:rte}) was checked
for the energy balance equation (\ref{eq:econs}), together with the surface
flux condition
\be
    4 \pi \int_0^{\infty} H_x ({m=0}) dx = 4 \pi    H_0 = \sigma_{\rm SB} T_{\rm eff}^4 .
\ee
The relative flux error 
\be
     \varepsilon_{\rm H}(m) = 1 - \frac{H_0}{\int_0^{\infty} H_x (m) dx},
\ee
and the energy balance error 
\be  \label{eq:econs1}
 \varepsilon_{\Lambda}(m) = \frac{1}{2}\! \int^{\infty}_{0} \!\! \! dx \!\! \int^{+1}_{-1} \!\! 
\left[\sigma(x,\mu) + k(x)\right]  \left[I(x,\mu)-S(x,\mu)\right]  d\mu
\ee
 were calculated as  functions of depth. 
Temperature corrections were then evaluated using three different procedures.  
In the upper atmospheric layers, we used the integral $\Lambda$-iteration method,
modified for Compton scattering, based on the energy balance equation (\ref{eq:econs}). 
The temperature correction for a particular depth was found as 
\be
     \Delta T_{\Lambda} = -  \varepsilon_{\Lambda}(m)  \ 
     \left(\int_0^{\infty}
 \left[ \frac{\Lambda_{\rm d}(x)-1}{1-\alpha(x)\Lambda_{\rm d}(x)} \right]
k(x)\, \frac{dB_x}{dT}\, dx  \right) ^{-1},
\ee  
where $\alpha(x)=\sigma_{\rm CS}(x)/(k(x)+\sigma_{\rm CS}(x))$, and
$\Lambda_{{\rm d}}(x)$  is the diagonal matrix element of the $\Lambda$-operator. 
Here $\sigma_{\rm CS}(x)$ is the Compton scattering opacity 
averaged over the relativistic Maxwellian electron distribution 
(see Eq. (A16) in \citealt{PS96}, which is  equivalent to Eq. (\ref{eq:scatopac}) 
if one ignores the induced scattering). 
In the deep layers, we used the Avrett-Krook flux correction based on the
relative flux error $\varepsilon_{\rm H}(m)$. 
Finally, the third procedure was the surface correction based on the emergent flux error
(see \citealt{Kurucz:70} for a detailed description of the methods).

The iteration procedure is repeated until the relative flux error is smaller than 0.1\%, and the relative flux 
derivative error is smaller
than 0.01\%.   As a result,  we obtain a self-consistent NS  model atmosphere, together with the emergent 
spectrum of radiation. 
We note that this accuracy is unachievable for luminous models with $g_{\rm rad} \approx g$, 
and that these models can have larger  relative flux errors, up to 2--3\%.

\subsection{Accuracy of computation}

To compute a new extended set of hot NS model atmospheres, we accelerated the convergence of the 
iterations 
of the radiation transfer equation by using the source function from the previous temperature iteration 
as the first approximation (see Appendix \ref{sec:method_rte}). 
However, in every fifth temperature iteration the radiation transfer equation was solved using the pure thermal 
source function as  a first approximation.  
We compared a pure hydrogen model atmosphere computed for 
$T_{\rm eff}= 1.8\times 10^7$ K and $\log g$ = 14.0 (the fiducial model) 
using this accelerated approach with the model computed without acceleration.
 The temperature structures differ by less than 0.3 \%, 
and the differences between the emergent spectra  are about 1\%  in the  3--20 keV energy range 
(typical of {\it RXTE}/PCA)  and larger in the Wien tail (Fig.~\ref{fig:1}).

As a convergence criterion  for the solution of the radiation transfer equation, we chose 
the maximum relative error of $10^{-4}$ in the mean intensity at all depths and energies. 
To determine the uncertainty in the final spectrum caused by this criterion, 
we compared the emergent spectra computed for the same model atmosphere with the accuracies
of $10^{-4}$ and $10^{-5}$ (see {\it top panel} of Fig.~\ref{fig:1}). 
We see that the relative error is smaller than 1\%  at all energies, which is then the intrinsic accuracy 
of our model spectra. 
We note that a similar error is introduced into the angular dependence of the specific intensities 
by ignoring polarization \citep[see e.g.][ compare his Tables XV and XXIV]{Cha60}.

\subsection{Various RFs and the Kompaneets operator}

\begin{figure*}
\begin{center}
\includegraphics[width= 0.73\columnwidth]{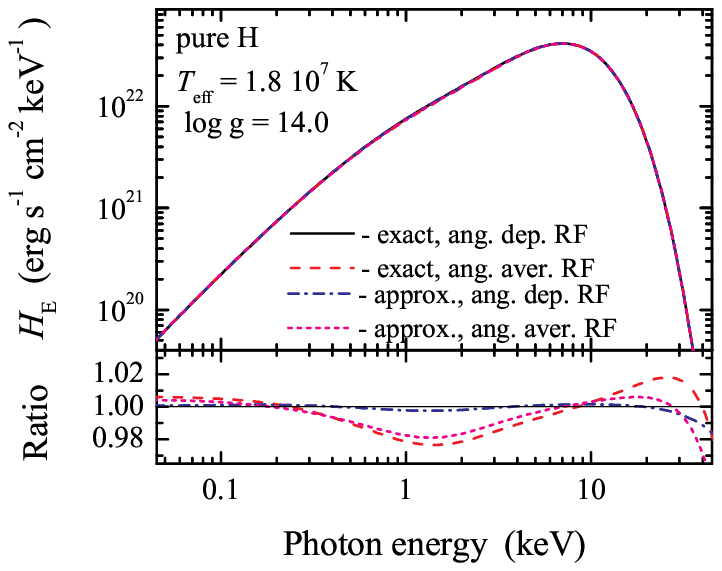}
\includegraphics[width= 0.73\columnwidth]{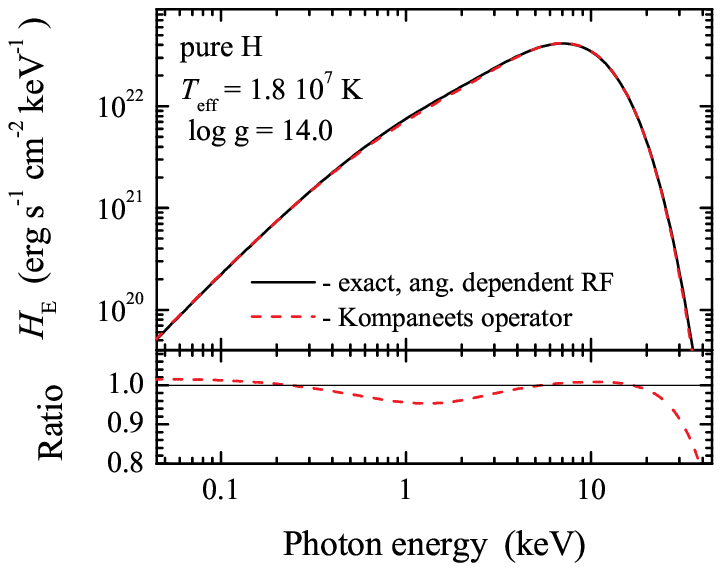}
\includegraphics[width= 0.73\columnwidth]{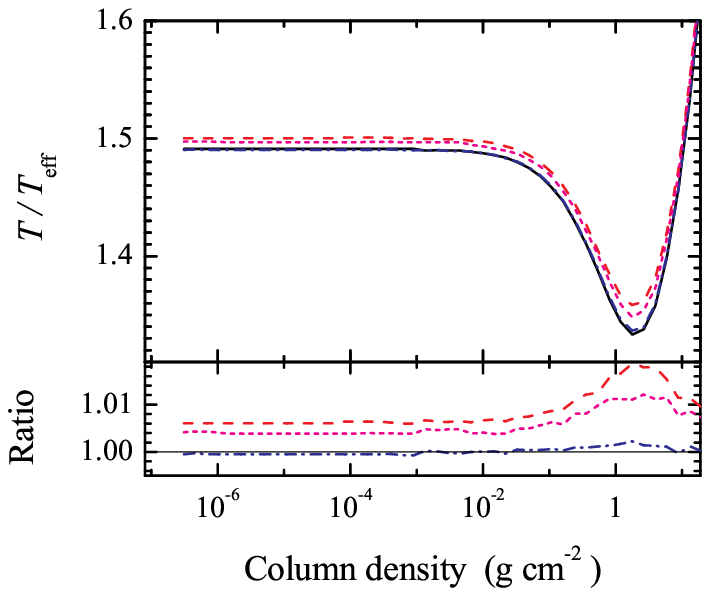}
\includegraphics[width= 0.73\columnwidth]{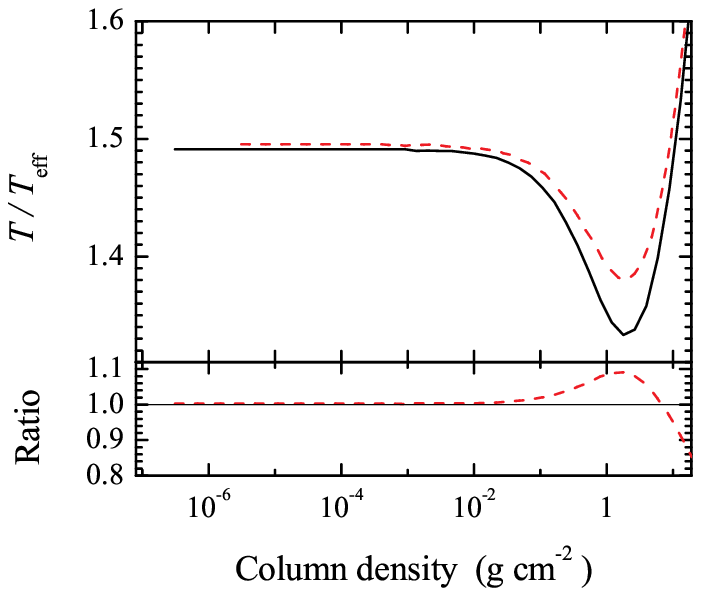}
\caption{\label{fig:2}
{\it Left panels} present the emergent spectrum ({\it top panel}) and the temperature structure ({\it bottom panel}) of the fiducial model 
computed using four different RFs: 
(a)  exact angle-dependent (solid curves), 
(b) exact angle-averaged  (dashed curves), 
(c)  approximate angle-dependent (dot-dashed curves),  and
(d) approximate angle-averaged (dotted curves). 
In the {\it top lower sub-panel}, the ratios of the spectra (b), (c), and (d)  to the spectrum for case (a)
 are shown. 
In the {\it bottom lower sub-panel}, the ratio of the temperature structures (b), (c), and (d)  to that of 
case (a) are shown. 
 {\it Right panels} present the emergent spectrum ({\it top panel}) and the temperature structure ({\it bottom panel}) 
of the fiducial model 
computed using an exact angle-dependent RF (solid curves)  and the Kompaneets operator (dashed curves). 
} 
\end{center} 
\end{figure*}

\begin{figure*}
\begin{center}
\includegraphics[width= 0.73\columnwidth]{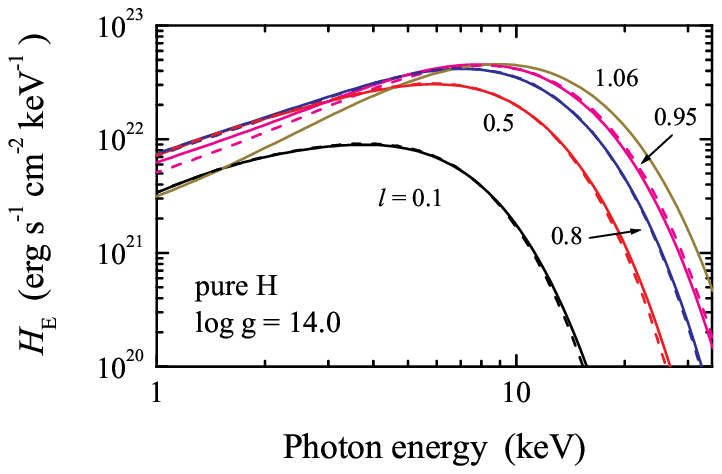}
\includegraphics[width= 0.73\columnwidth]{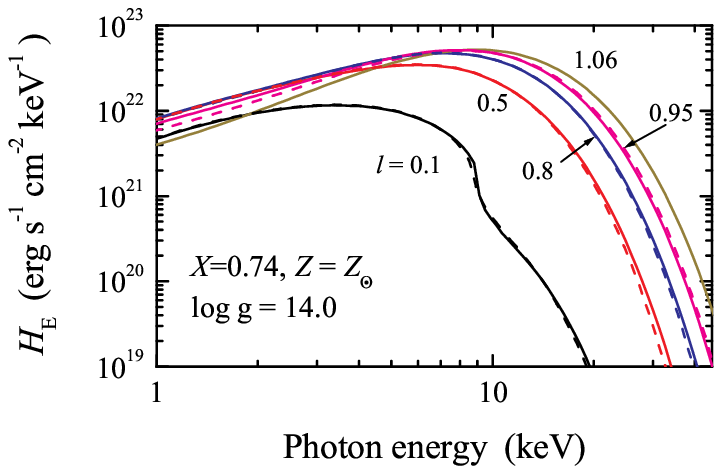}
\includegraphics[width= 0.73\columnwidth]{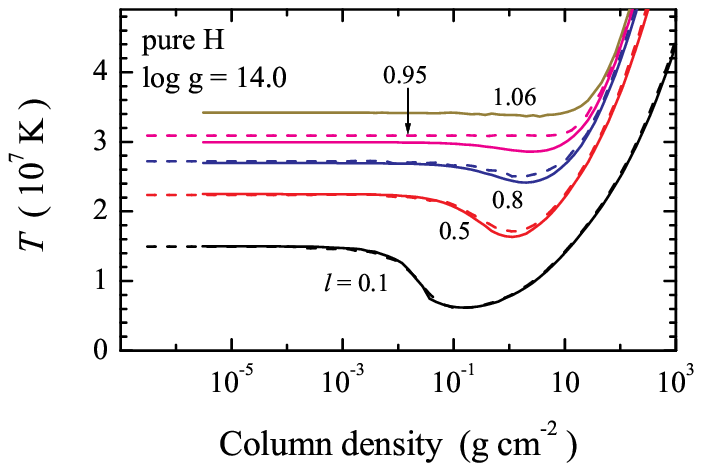}
\includegraphics[width= 0.73\columnwidth]{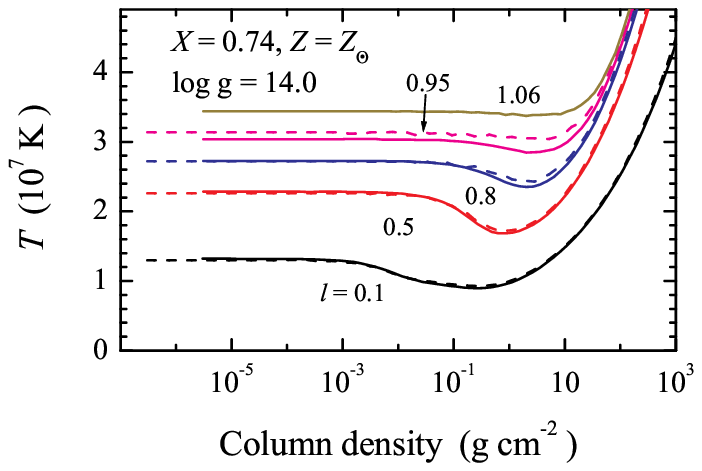}
\caption{\label{fig:3}
Comparison of  the emergent spectra  ({\it top panel}) and the temperature structures  ({\it bottom panel}) computed using the 
present code with the exact RF (solid curves) and  the previous code  employing the Kompaneets operator (dashed curves).
The models for various relative luminosities are marked by corresponding numbers next to the curves. 
{\it Left panels} are for pure H  and {\it right panels} are for the solar abundances.   
} 
\end{center} 
\end{figure*}

A comparison of  model atmospheres, which were computed using our new code with four different RFs (exact 
angle-dependent, exact angle-averaged,  approximate  angle-dependent,  and approximate  angle-averaged) 
is shown in the left panels of Fig.~\ref{fig:2}.  
The model computed using the approximate angle-dependent RF is almost indistinguishable from 
 the reference model calculated using the exact  angle-dependent RF. 
This is unsurprising, because the approximate function matches the accurate one very well up to the 
temperatures of about 100 keV. 
Similar results were obtained by \citet{PS96} for the Comptonization spectra in the optically thin slabs. 
The differences are smaller than 1\% for the emergent spectra and below 0.3\% for the temperature structure. 
Deviations of model atmospheres computed using the angle-averaged RF 
of the reference model are more significant, 
at about 2\% for both  the temperature structure and spectra. 

The model atmosphere calculated employing the Kompaneets operator has more significant differences 
from the reference model (see right panels in Fig.~\ref{fig:2}), of
up to 10\% in the temperature structure and about 3--4\% in the emergent flux at 1 keV. 
The deviations are much smaller ($< 1\%$) near the spectral maximum  and larger in the Wien tail. 
A rather good agreement between model atmospheres computed with the relativistic exact angle-dependent RF 
and  the non-relativistic angle-independent Kompaneets operator again is unsurprising, 
because temperatures of the upper atmosphere 
layers where the emergent spectra form are sufficiently low ($\sim$2--4 keV) and relativistic corrections are small.

\section{New grid of models}
\label{s:newgrid}
 
\subsection{General properties}

We computed a new set of hot NS model atmospheres using the exact 
relativistic angle-dependent RF. 
The models were calculated for six chemical compositions (pure hydrogen, pure helium, and solar
 hydrogen/helium mix with various heavy element abundances: of solar and 0.3, 0.1 and 0.01 of solar). 
For every chemical composition, 26--28 models with  relative luminosities 
spanning the interval from $l= 0.001$ to 1.06--1.10  for three values of the surface gravity 
($\log g = $14.0, 14.3 and 14.6) were calculated. 
Because the radiative acceleration in our models is smaller than in the models based on the Thomson opacity 
owing to the Klein-Nishina reduction in the cross-section (see below), 
there exist formally ``super-Eddington'' (relative to $\leddth$) models. 
The Klein-Nishina reduction depends on the electron temperature, which is higher for larger surface gravities, 
therefore the limiting luminosity is higher for larger $\log g$.  
 
\begin{figure}
\begin{center}
\includegraphics[width= 0.73\columnwidth]{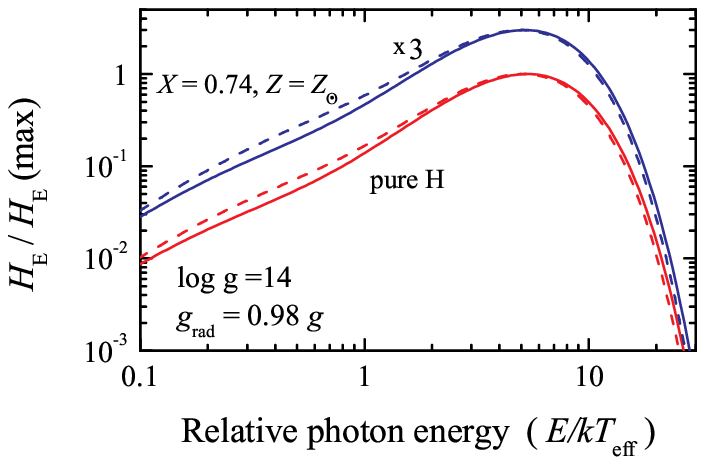}
\includegraphics[width= 0.73\columnwidth]{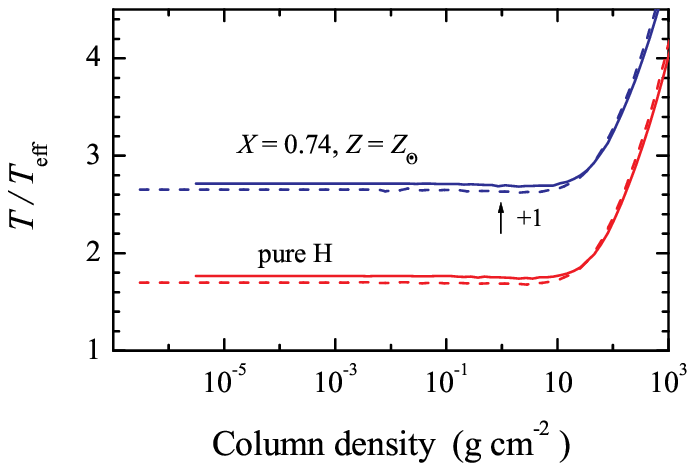}
\caption{\label{fig:4}
Comparison of the emergent spectra ({\it top panel}) and the temperature structures  
({\it bottom panel}) of the new (solid curves) and the 
old (dashed curves) models  with the same $g_{\rm rad}/g$. The models
 are computed for pure hydrogen as well as solar abundance. 
The models have different effective temperatures, 
therefore the spectra are normalized to the maximum flux  and  plotted against the scaled photon energy.  
For clarity, the spectra of solar abundance models are shifted by a factor of three, and the relative 
temperatures are shifted by adding one.} 
\end{center} 
\end{figure}

\begin{figure}
\begin{center}
\includegraphics[width= 0.73\columnwidth]{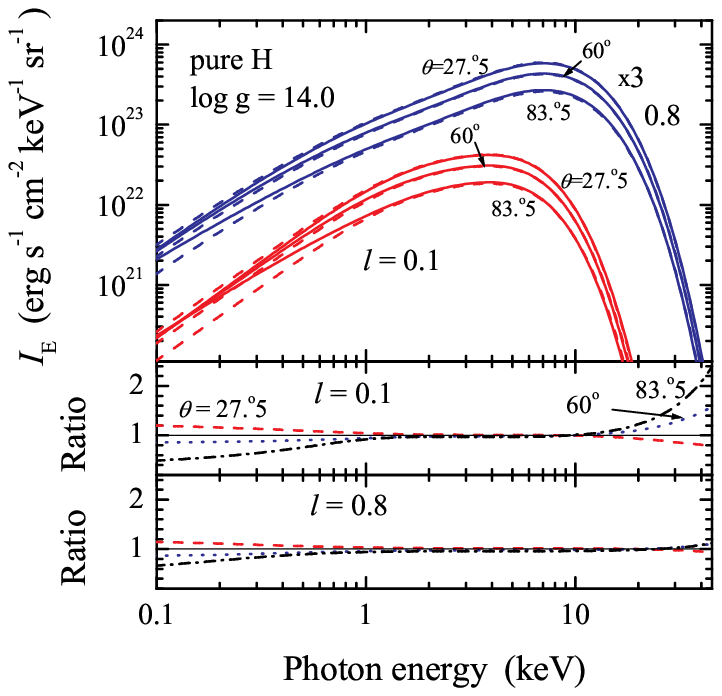}
\includegraphics[width= 0.73\columnwidth]{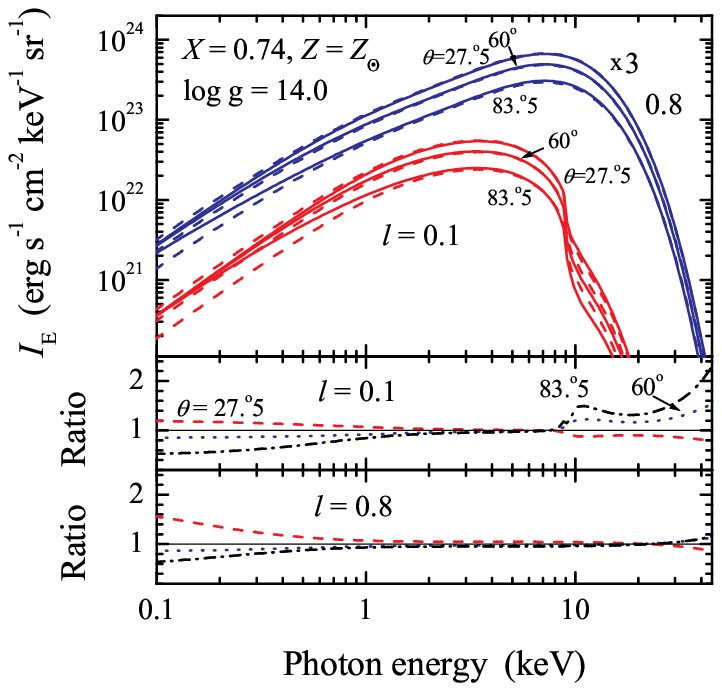}
\caption{\label{fig:5}
Comparison of the emergent specific intensity for pure hydrogen ({\it top panel})  and  solar abundance models  ({\it bottom panel})  
with the electron-scattering limb-darkening law (dashed curves) for two relative luminosities of $l=0.1$ and 0.8. 
The $l=0.8$ models are multiplied by a factor of three for clarity. 
The lower sub-panels show the ratios of the corresponding models. 
} 
\end{center} 
\end{figure}

Examples of emergent spectra and temperature structures for the models with $\log g$ = 14.0 and various 
chemical compositions (pure hydrogen and solar mix) are shown in Fig.~\ref{fig:3}.\footnote{The spectral 
energy distributions of fluxes and specific intensities 
for all models are described in Appendix \ref{app:D}.}
The corresponding emergent spectra and temperature structures computed with the old code employing the 
Kompaneets operator are also shown. 
At low luminosities, the models are very close to each other, which is 
expected as at low temperatures the diffusion  (Kompaneets) approximation is an accurate representation 
of Compton scattering. 
For models close to the Eddington limit, the treatment of the radiative acceleration becomes important.  
Contribution of the radiation force to the hydrostatic equilibrium $g_{\rm rad}/g$ is smaller in the new models 
despite $l$ being the same for both model sets. 
It is well-known that the model spectra with large contributions of radiative acceleration are harder, and their
 surface temperatures are higher \citep{Londonetal:86,Lapidus:86,Ebisuzaki87,Pavlov.etal:91}. 
In complete agreement with this, the old models with large $l$ are harder and hotter than new models.  
However,  for the same $g_{\rm rad}/g$ the new model spectra  are hotter. 
Normalizing the spectra to the maximum flux and plotting them against the scaled photon energies 
$E/kT_{\rm eff}$, one also sees that the new spectra are harder  (see Fig.~\ref{fig:4}).

\subsection{Limb-darkening}

\begin{figure}
\begin{center}
\includegraphics[width= 0.73\columnwidth]{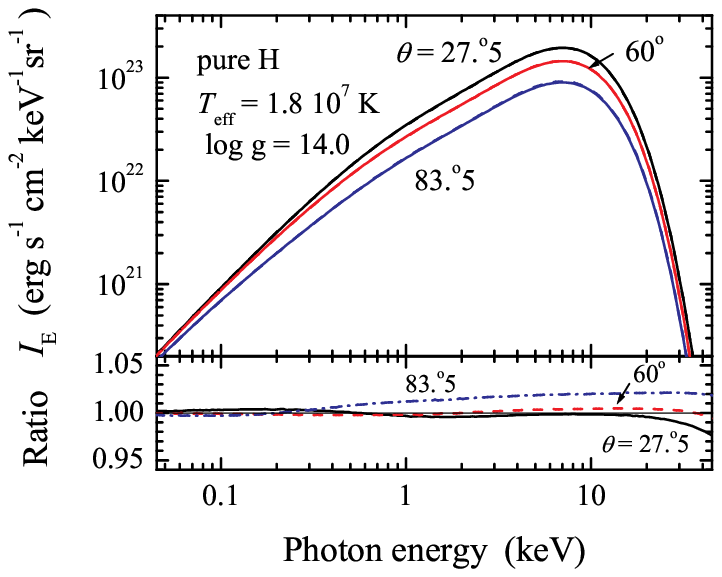}
\includegraphics[width= 0.73\columnwidth]{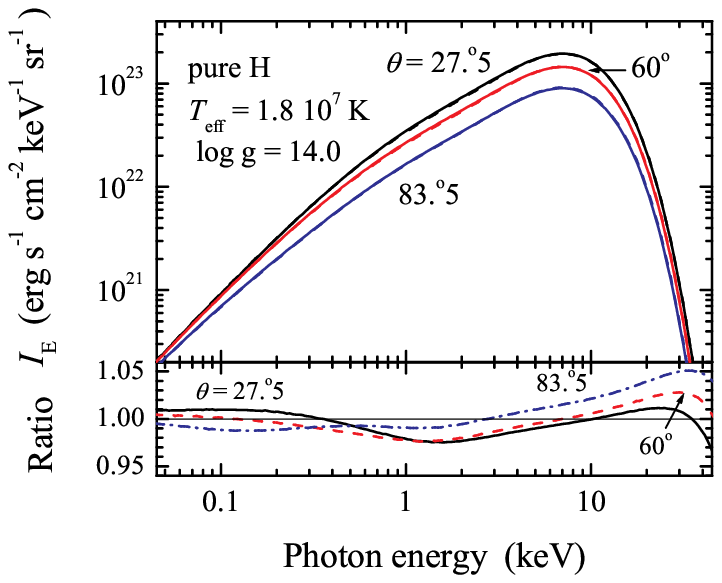}
\caption{\label{fig:6}
Comparison of the emergent specific intensity at three angles 
(from the Gaussian quadrature $\mu=0.113, 0.5, 0.887$, i.e. $\theta =  83\fdg5, 60\degr, 27\fdg5$) 
for the fiducial model  computed using the exact angle-dependent RF (solid curves in both panels) 
with the spectra based on the approximate angle-dependent RF ({\it top panel}, dashed curves) and 
exact angle-averaged RF ({\it bottom panel}, dashed curves). 
In the  bottom sub-panels,
 the ratios of the accurate spectra to the approximate spectra  for the three angles are shown. 
} 
\end{center} 
\end{figure}

Knowledge of the angular distribution of the emergent radiation is important when only part 
of the star is visible (for example, is partially blocked by the accretion disk), or when there are
 inhomogeneities at the NS surface related, 
for example, to the varying gravitational acceleration due to the rapid rotation. Computation of the amplitude of 
reflection from the accretion disk also requires that information.
The limb-darkening law (i.e. angular dependence of the intensity) 
for the radiation emerging from the optically thick electron-scattering-dominated atmosphere 
is described by the $H^{(0)}(\mu)$ function \citep{ChaBre47,Cha60,sob49,Sob63}. 
A simple approximation to this function is $I(\mu) \approx 1+2.06\mu$.  
Our simulations show that the intensity closely  (within a couple of per cents) follows the $H^{(0)}(\mu)$ 
function  around 
the peak of the spectrum, in the observed energy range (see  Fig.~\ref{fig:5}). 
At photon energies below 1 keV, the computed angular distribution becomes more isotropic, 
because there free-free absorption dominates over
 electron scattering and the upper atmosphere layers are almost isothermal. 
The low-inclination  intensity is  above and the  high-inclination intensity is  below 
the electron-scattering limb-darkening law at energies above the peak, because 
the temperature of the layer where the photons originate drops with the inclination. 
The iron absorption edge at $\sim$9 keV significantly  affects 
the angular distribution for the solar abundance models (Fig.~\ref{fig:5}, {\it bottom panel}). 
Above the edge, radiation becomes more directed along the surface normal.

The angular distribution of the intensity also depends on the specific form of the RF used in the calculations. 
The approximate angle-dependent gives  results very close to the reference intensity spectra,  
within 2\% for the largest angle $\theta$ (see Fig.~\ref{fig:6}, {\it top panel}).
The exact, but angle-averaged, RF, does not give such accurate results (see Fig.~\ref{fig:6}, {\it bottom panel}).

\subsection{Color correction factors}

\begin{figure}
\begin{center}
\includegraphics[width= 0.73\columnwidth]{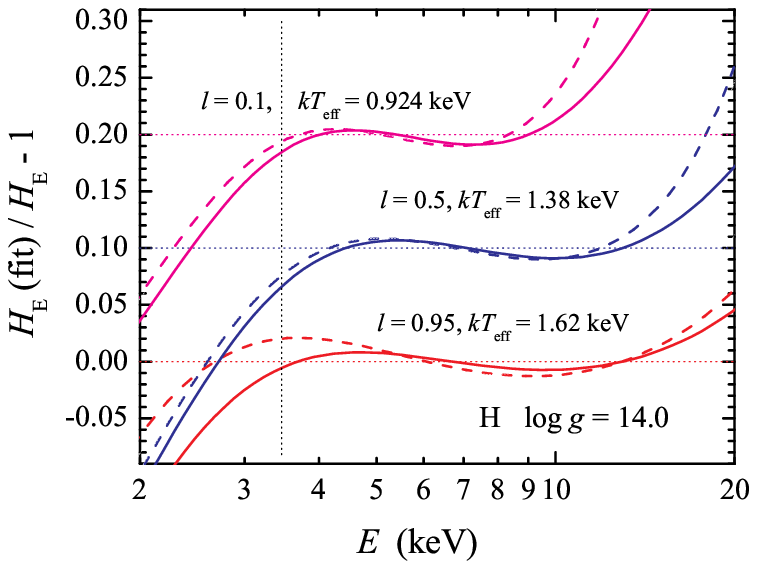}
\includegraphics[width= 0.73\columnwidth]{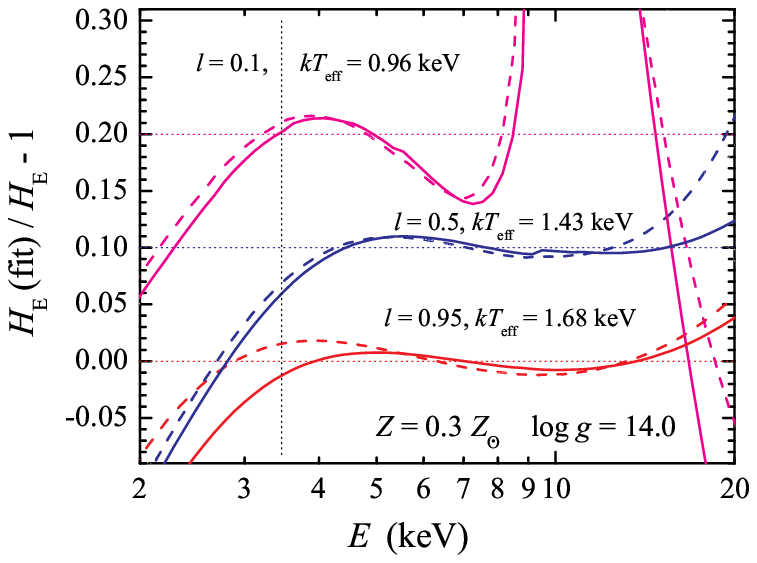}
\caption{\label{fig:7}
Relative deviations of the new, exact RF-based (solid curves)  and old Kompaneets-based (dashed curves)   
spectra from the best-fit diluted blackbodies versus photon energy for hydrogen ({\it top panel}) 
and solar H/He mixture with $Z=0.3Z_{\odot}$  ({\it bottom panel}) low gravity ($\log g$ =14.0) models. 
Corresponding relative luminosities and the effective temperatures are given at the curves. 
The vertical dotted line shows the lower boundary of the energy band, where the fitting procedures were performed. 
For clarity, models with $l$ = 0.1 and 0.5 are shifted up by 0.2 and 0.1, respectively.  
} 
\end{center} 
\end{figure}

\begin{figure}
\begin{center}
\includegraphics[width= 0.73\columnwidth]{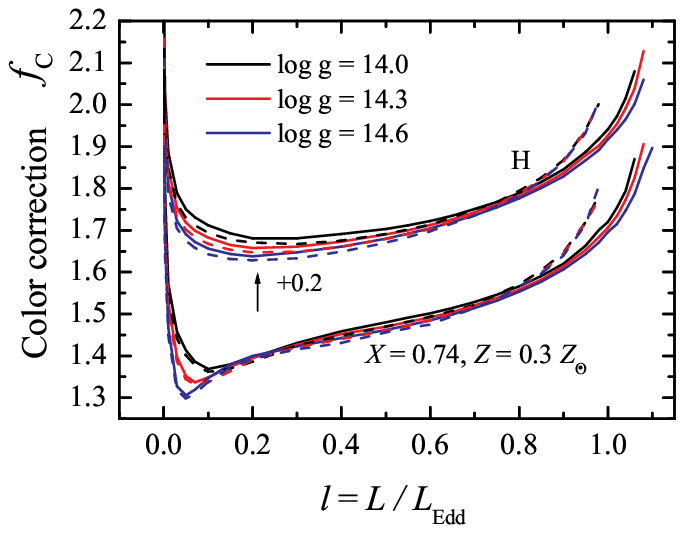}
\includegraphics[width= 0.73\columnwidth]{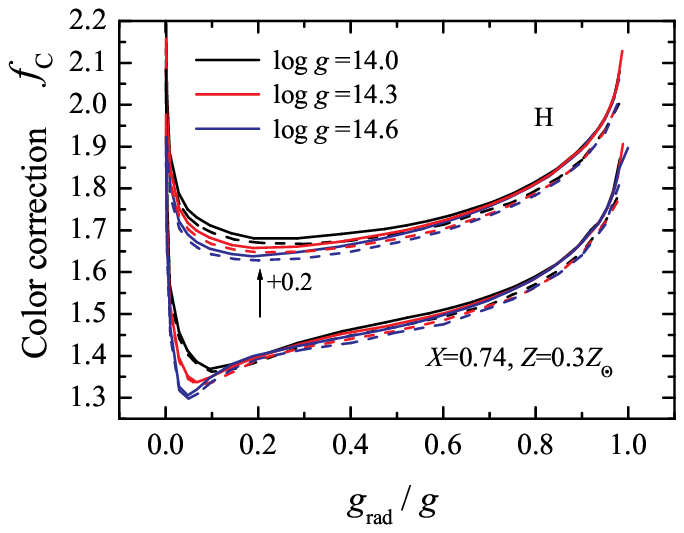}
\caption{\label{fig:8}
Color correction factors $f_{c,1}$ (computed using method 1 from SPW11) 
for model atmospheres of two chemical compositions 
(pure hydrogen and solar hydrogen/helium mix with 30\% of solar heavy-element abundance)
and different $\log g$ as functions of the relative luminosity  $l$   ({\it top panel}) or  $g_{\rm rad}/g$ ({\it bottom panel}).  
The new models based on the exact RF are shown by the solid curves, and the old models, 
based on the Kompaneets operator, by dashed curves. 
} 
\end{center} 
\end{figure}

All computed emergent spectra of the new atmosphere models were fitted by a diluted  blackbody spectrum
\be \label{eq:fit}
      F_{E} \approx w B_{E}(\fc T_{\rm eff})
\ee
using five different fitting procedures described in SPW11. 
We calculated the color correction $\fc$ and dilution $w$ factors
in the  energy band (3--20)$\times(1+z)$ keV corresponding to the observed range of
 the {\it RXTE}/PCA detector.
We calculated redshifts from $\log g$  by adopting a NS mass equal to
1.4$M_{\odot}$ (see Eqs.~(\ref{eq:g_def}) and (\ref{eq:redshift_def})): 
for $\log g$ = 14.0, 14.3, and 14.6, we get $R$ = 14.80, 10.88, 8.16 km  
and  $z$ = 0.18, 0.27, 0.42, respectively.
Varying the mass in the interval 1--2$M_{\odot}$ has a smaller  than 0.1\% effect on the color corrections (SPW11).
The results of the fitting procedures are presented in Table~\ref{tab:fc} (see also Appendix \ref{app:D}).  

\begin{figure}
\begin{center}
\includegraphics[width= 0.73\columnwidth]{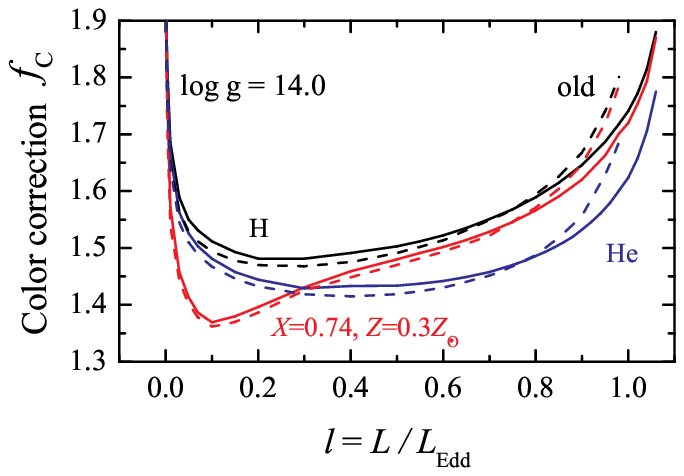}
\includegraphics[width= 0.73\columnwidth]{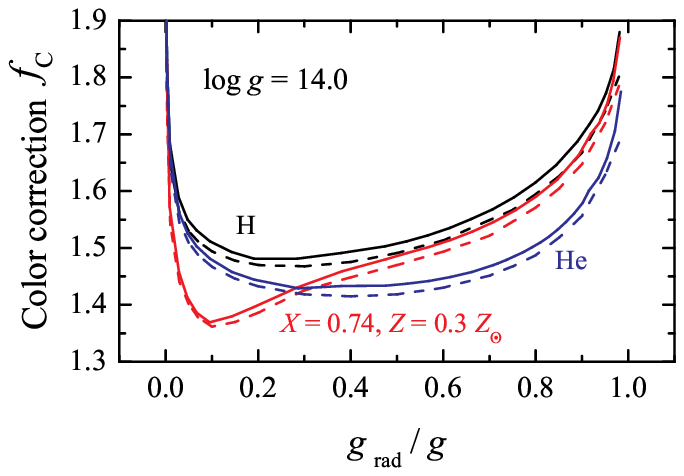}
\caption{\label{fig:9}
Same as Fig.~\ref{fig:8}, but for different chemical compositions  and $\log g=14.0$.   
} 
\end{center} 
\end{figure}

\begin{figure}
\begin{center}
\includegraphics[width= 0.73\columnwidth]{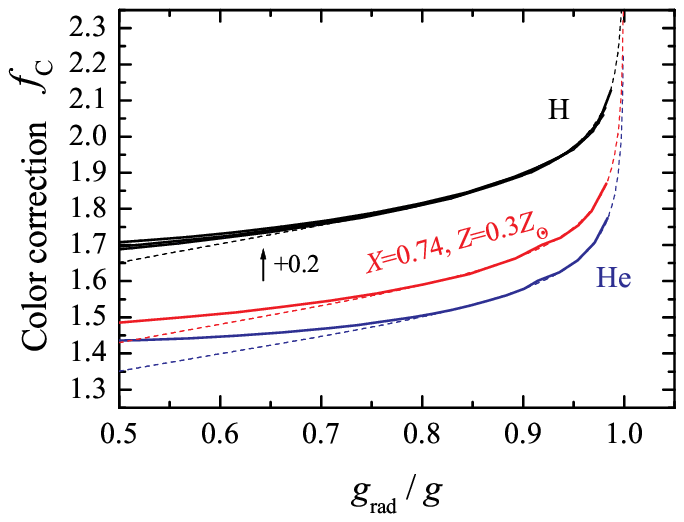}
\caption{\label{fig:10}
Color correction factor $\fc$ versus $g_{\rm rad}/g$ for models of various chemical compositions. 
Solid curves are the results of calculations based on the exact Compton RF, 
the dotted curves give the approximation (\ref{eq:fcfit}).  
For clarity, the curves for pure H models  (shown for three $\log g$) are shifted up by +0.2. 
} 
\end{center} 
\end{figure}

Deviations of the new model spectra from the best-fit diluted blackbodies are similar 
to those for the old spectra based on the Kompaneets operator 
(Fig.~\ref{fig:7}). 
Comparison of the new and old color correction factors is shown in Figs.~\ref{fig:8} and \ref{fig:9}.  
Although the old $\fc$--$l$ dependences for different gravities were almost identical at high luminosities, 
we see that now these dependences deviate, because of the dependence of the limiting 
relative luminosity on  $\log g$.
However, the old and new color-correction factors have very similar dependences on
 $g_{\rm rad}/g$ for all $\log g$ 
and  chemical compositions (see Figs.~\ref{fig:8} and \ref{fig:9}), with the new $\fc$ being about 1\% larger. 
The difference grows at $g_{\rm rad}/g$ close to unity. 

\citet{Pavlov.etal:91} derived an approximate analytical formula for the ratio of the surface model 
atmosphere temperature to the effective one
 \be \label{eq:tsurfteff}
\frac{ T_{\rm surf}}{T_{\rm eff}} \approx \left( 0.14 \ln \frac{3+5X}{1-l} + 0.59\right) ^{-4/5} 
\left(  \frac{3+5X}{1-l} \right)^{2/15} l^{\,3/20} ,
 \ee
which is correct for the highly luminous ($l>0.9$) model atmospheres. 
The numerical constants 0.14 and 0.59 were found from the fitting of their 
model atmospheres based on the Kompaneets operator description 
of Compton scattering.  
If the models were instead computed using an exact RF for Compton scattering, it is obvious 
that $l$ should be substituted by the relative radiation acceleration $g_{\rm rad}/g$.  
We fit our color correction factors computed using the first fitting procedure (see SPW11) with a
 formula similar to Eq. (\ref{eq:tsurfteff}) 
and found numerical constants that also  depend  on the chemical composition
\bea \label{eq:fcfit}
\fc &\approx& \left( \left[ 0.102+0.008X \right] \ln \frac{3+5X}{1-g_{\rm rad}/g} + 0.63-0.06X \right)
 ^{-4/5} \nonumber \\  
&\times& \left(  \frac{3+5X}{1-g_{\rm rad}/g} \right) ^{2/15} (g_{\rm rad}/g )^{3/20}.
 \eea
This approximation works well for  $g_{\rm rad}/g > 0.8$ (see Fig.~\ref{fig:10}). 

\begin{table*}
\caption{\label{tab:fc}
Color-correction and dilution factors from the blackbody fits to the spectra of hydrogen 
atmosphere  models at $\log g=14.0$. }
\begin{center}
\begin{tabular}{cccccccccccc}
\hline 
\hline \multicolumn{11}{c}{  $X=1$ $\quad$ $Z=0$  $\,\,\,\, \log g = 14.0 \,\,\,\, \teddth= 1.644$ keV  $\quad R=   14.80$ km  $\quad  z=   0.18$} \\
Set &   $l$     & $g_{\rm rad}/g$   &   $T_{\rm eff}$ (keV)   &   $f_{\rm c,1}$ &  $f_{\rm c,2}$    & $f_{\rm c,3}$ &       
    $f_{\rm c,4}$ & $f_{\rm c,5}$ & $w_1 f_{\rm c,1}^4$  & $w_2 f_{\rm c,2}^4$ &  $w_3 f_{\rm c,3} ^4 $  \\
\hline
 1 & 0.001 & 0.001 & 0.292 & 2.015 & 2.009 & 2.018 & 1.806 & 1.571 & 0.698 & 0.704 & 0.695 \\
 1 & 0.003 & 0.003 & 0.385 & 1.830 & 1.826 & 1.833 & 1.662 & 1.473 & 0.796 & 0.800 & 0.794 \\
 1 & 0.010 & 0.009 & 0.520 & 1.686 & 1.682 & 1.687 & 1.553 & 1.382 & 0.877 & 0.878 & 0.876\\
 1 & 0.030 & 0.028 & 0.684 & 1.589 & 1.587 & 1.589 & 1.525 & 1.375 & 0.926 & 0.926 & 0.926 \\
 1 & 0.050 & 0.048 & 0.777 & 1.550 & 1.550 & 1.550 & 1.531 & 1.407 & 0.942 & 0.942 & 0.942 \\
 1 & 0.070 & 0.067 & 0.845 & 1.531 & 1.530 & 1.530 & 1.533 & 1.431 & 0.949 & 0.949 & 0.949 \\
 1 & 0.100 & 0.096 & 0.924 & 1.512 & 1.511 & 1.511 & 1.528 & 1.451 & 0.954 & 0.954 & 0.954 \\
 1 & 0.150 & 0.143 & 1.023 & 1.493 & 1.492 & 1.492 & 1.512 & 1.462 & 0.963 & 0.962 & 0.963 \\
 1 & 0.200 & 0.190 & 1.099 & 1.481 & 1.479 & 1.481 & 1.505 & 1.464 & 0.960 & 0.959 & 0.960 \\
 1 & 0.300 & 0.285 & 1.217 & 1.481 & 1.476 & 1.480 & 1.500 & 1.478 & 0.971 & 0.969 & 0.971 \\
  1&   0.400  & 0.380 &  1.307  &  1.491  &  1.486  &  1.491  & 1.511  & 1.500  &  0.972  &  0.970  &  0.972   \\ 
  1&   0.500  & 0.474 &  1.382  &  1.503  &  1.497  &  1.503  & 1.521  & 1.515  &  0.974  &  0.972  &  0.974  \\ 
  1&   0.550  & 0.521 &  1.416  &  1.512  &  1.506  &  1.512  & 1.530  & 1.526  &  0.975  &  0.972  &  0.975 \\  
  1&   0.600  & 0.567 &  1.447  &  1.522  &  1.516  &  1.522  & 1.539  & 1.538  &  0.977  &  0.973  &  0.976  \\
  1&   0.650  & 0.614 &  1.476  &  1.535  &  1.530  &  1.535  & 1.551  & 1.553  &  0.979  &  0.976  &  0.979  \\
  1&   0.700  & 0.661 &  1.504  &  1.551  &  1.546  &  1.551  & 1.566  & 1.569  &  0.981  &  0.977  &  0.980   \\
  1&   0.730  & 0.707 &  1.530  &  1.568  &  1.563  &  1.567  & 1.583  & 1.586  &  0.982  &  0.979  &  0.982  \\      
  1&   0.800  & 0.733 &  1.555  &  1.589  &  1.586  &  1.588  & 1.600  & 1.608  &  0.985  &  0.983  &  0.984  \\     
  1&   0.850  & 0.799 &  1.578  &  1.615  &  1.613  &  1.613  & 1.624  & 1.634  &  0.987  &  0.986  &  0.987  \\
  1&   0.900  & 0.844 &  1.601  &  1.646  &  1.648  &  1.644  & 1.653  & 1.665  &  0.992  &  0.993  &  0.991  \\
  1&   0.950  & 0.890 &  1.623  &  1.687  &  1.694  &  1.682  & 1.690  & 1.705  &  0.997  &  1.001  &  0.996  \\ 
  1&   0.980  & 0.917 &  1.635  &  1.718  &  1.729  &  1.712  & 1.718  & 1.735  &  1.000  &  1.007  &  0.999   \\ 
  1&   1.000  & 0.934 &  1.644  &  1.741  &  1.756  &  1.733  & 1.739  & 1.757  &  1.002  &  1.011  &  1.000  \\ 
  1&   1.020  & 0.952 &  1.652  &  1.772  &  1.793  &  1.762  & 1.767  & 1.787  &  1.006  &  1.018  &  1.003  \\
  1&   1.040  & 0.969 &  1.660  &  1.815  &  1.844  &  1.803  & 1.807  & 1.828  &  1.010  &  1.026  &  1.007  \\
  1&   1.060  & 0.981 &  1.668  &  1.880  &  1.921  &  1.864  & 1.869  & 1.888  &  1.012  &  1.036  &  1.009  \\ 
   \hline
\end{tabular}
\end{center}
\tablefoot{Results for other chemical compositions and gravities  are given in Table D.1 available in electronic form at 
the Centre de Donn\'ees astronomiques de Strasbourg (CDS). }
\end{table*}

\subsection{Radiative acceleration}

\begin{figure}
\begin{center}
\includegraphics[width= 0.73\columnwidth]{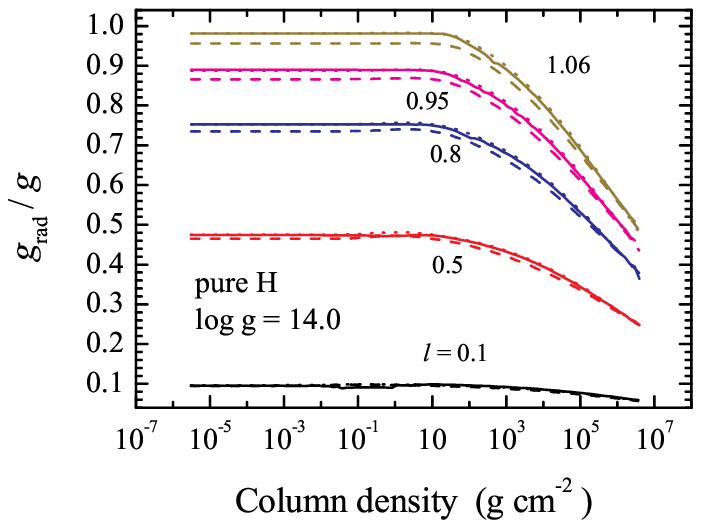}
\includegraphics[width= 0.73\columnwidth]{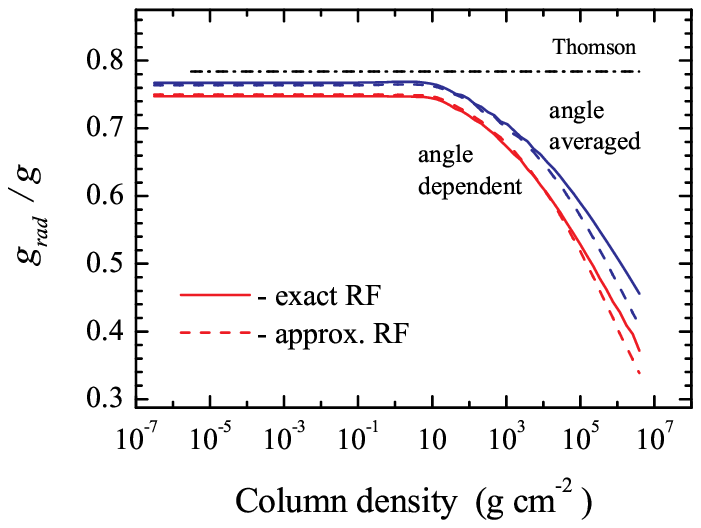}
\caption{\label{fig:11}
{\it Top panel:} Comparison of the relative radiative acceleration in pure hydrogen atmospheres 
(solid curves) with 
our approximation given by Eq. (\ref{eq:rossel}) (dotted curves, which nearly coincide with solid curves) 
and with Paczynski's approximation (\ref{eq:pacz})  (dashed curves). 
{\it  Bottom panel:} Comparison of the relative radiative acceleration for 
the fiducial  model computed with different RFs:
exact and approximate, angle-dependent, and angle-averaged.  
The relative radiative acceleration for Thomson scattering is also shown. }
\end{center} 
\end{figure}

The radiative acceleration can formally be  represented as a product of the flux and 
the temperature-dependent effective opacity
\be \label{eq:grad_opac}
g_{\rm rad} = \kappa (T)  \frac{\sigma_{\rm SB}T^4_{\rm eff}}{c}.
\ee
This expression can alternatively be written as 
\be \label{eq:grad_lum}
\frac{g_{\rm rad}}{g} = l\ \frac{\kappa (T)}{\kappae} . 
\ee
In diffusion approximation, $\kappa (T)$ is given by the Rosseland mean opacity. 
When  electron scattering dominates, it is often approximated (neglecting  the electron degeneracy)  
as \citep{Pacz:83}
\be 
\label{eq:pacz}
\kappa (T) = \kappaR (T) \approx \kappae 
\left[1+\left(\frac{kT}{38.8\ {\rm keV}} \right)^{0.86} \right]^{-1}  .
\ee  
This approximation  is based on calculations by \citet{BY76}, who underestimated the opacity 
at low temperatures, where certain approximations were made because of severe numerical problems.
A better approximation in the range 2--50 keV that is of interest here is (J. Poutanen et al., in preparation) 
\be 
\label{eq:rossel}
\kappaR (T) \approx \kappae \left[1+\left(\frac{kT}{39.4\ {\rm keV}} \right)^{0.976} \right]^{-1}  .
\ee  
It is clear that the radiative acceleration decreases in the deep, hotter atmosphere layers and  
the ratio of radiation pressure force  to the surface gravity decreases inwards (see Fig.~\ref{fig:11}).  
The radiative acceleration  is smaller than that corresponding to the Thomson opacity,  even in the upper atmosphere layers.  
The actual radiative acceleration in the surface layers (see {\it top panel} in Fig.~\ref{fig:11}) 
computed from the models using Eq.~(\ref{eq:grad})  is perfectly described by  Eqs.~(\ref{eq:rossel}) and (\ref{eq:grad_lum}) 
throughout the whole atmosphere (compare dotted and solid curves at the {\it top panel} in Fig.~\ref{fig:11}), while 
Paczynski's approximation underestimates it.

Radiative acceleration computed using the angle-averaged RFs is larger than that computed using 
angle-dependent RFs  (see  {\it bottom panel}   in Fig.~\ref{fig:11}), 
because the high-energy photons scatter on the relatively cold electrons in a predominantly forward 
direction, reducing the magnitude of the momentum transfer in comparison with the isotropic case. 
 
\begin{figure}
\begin{center}
\includegraphics[width= 0.73\columnwidth]{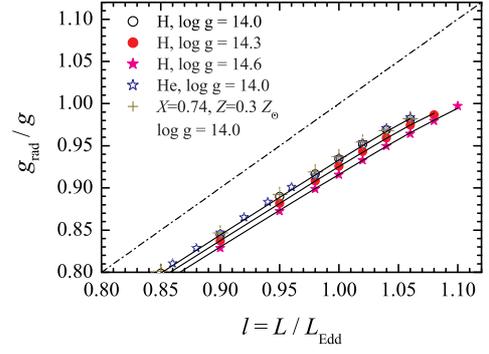}
\caption{\label{fig:12}
Dependence of the relative radiative acceleration $g_{\rm rad}/g$ on $l$. 
Symbols represent the results of calculations based on the exact Compton RF, 
the solid curves give the approximation  (\ref{eq:grfit}). 
The dot-dashed curve is  the $l=g_{\rm rad}/g$ relation.
} 
\end{center} 
\end{figure}

Radiative acceleration in the surface layers that we obtain from the models 
can be expressed through $T_{\rm eff}$ and $\fc$ as 
\be \label{eq:grfit}
\kappa (T=\fc T_{\rm eff}) = \kappae\  
\left[1+\left(\frac{kT}{38.8\ {\rm keV}} \right)^{\alpha_{\rm g}} \right]^{-1}, 
\ee
where $\alpha_{\rm g} = 1.01 + 0.067 (\log g -14.0)$ (see Fig.~\ref{fig:12}). 
The relation $g_{\rm rad}/g$--$l$  also slightly depends on the chemical composition, 
but the dependence on the surface gravity is stronger.

The relation  of the effective temperature to the effective opacity given by Eq. (\ref{eq:grfit})
allows us to estimate $g_{\rm rad}/g$ for the given model parameters $(l,T_{\rm eff})$ 
without actually computing the atmosphere models. 
We can make a first guess of $\fc$, substitute $T=\fc T_{\rm eff}$ to Eq. (\ref{eq:grfit}), 
then use this $\kappa$ in Eq.~(\ref{eq:grad_lum}), find a new $\fc$ through Eq.~(\ref{eq:fcfit}), and iterate.

\section{Application to observations}
\label{s:application}

In our previous works (S11 and SPW11), we suggested a new method for the 
determination of  NS radii and masses. 
It is based on the spectral (blackbody) normalization 
$K$ at late phases of PRE X-ray bursts depending on the color correction factor only 
\be \label{eq:Knorm}
   K \equiv \left( \frac{R_{\rm bb}\,{\rm (km)}}{D_{10}} \right) ^2 = \frac{1}{\fc ^4} \left( \frac{R\,{\rm (km)}
   \, (1+z) }{D_{10}}\right) ^2 = (A \fc)^{-4} ,
\ee
where $D_{10} = D/10\,{\rm kpc}$ is the distance. 
The observed relation $K^{-1/4}$--$F_{\rm bb}$ between the blackbody flux $F_{\rm bb}$ 
and normalization $K$   
can be fitted by the theoretical dependence $\fc$--$l$ obtained from the atmosphere models. 
The fit gives  two parameters: the value of $A=\left[R\,{\rm (km)}\, (1+z)/D_{10}\right]^{-1/2}$ and 
the observed Eddington flux  $\feddth = \leddth (1+z)^{-2} /(4\pi D^2)$.  
The distance-dependent quantities $A$ and $\feddth$ can be combined to the 
distance-independent Eddington temperature 
which is the apparent effective temperature corresponding to $\leddth$
\be \label{eq:teddinf}
\teddinf=  \left( \frac{gc}{\sigma_{\rm SB} \kappae } \right) ^{1/4} \frac{1}{1+z} = 6.4\times 10^9\ A\  
\feddth ^{1/4} \ \mbox{K}.
\ee
If the distance to the source is known (for example, for sources in globular clusters), we can 
plot three curves on the $M$ -- $R$ plane corresponding  to the values of $A$, $\feddth$, and $\teddinf$. 
Two crossing points give two pairs of NS mass and radius values that  satisfy  the observed data. 

It is now important to understand how the new models affect results based on formulae that use the
Thomson cross-section for electron scattering. We have seen that owing to the Klein-Nishina reduction in the 
cross-section, the actual Eddington limit is reached at luminosities of 6--10\% higher than $\leddth$ 
and the shape of the $\fc$--$l$ relation differs somewhat from the Kompaneets-based results. 
As an illustration, we consider  a long PRE X-ray burst of 4U\,1724--307 in the globular cluster Terzan~2 
 studied by S11, who obtained a rather large NS radius for this source ($R\ge 14$\, km). 
We now use new sets of  theoretical relations $\fc$--$l$ that we fit to the observed relation 
$K^{-1/4}$--$F_{\rm bb}$, taking pure hydrogen models as an example. 
We note that the new relations significantly depend on the surface gravity (see Fig.~\ref{fig:8}, {\it top panel}), 
therefore, in principle, it would be possible also to find  the surface gravity that provides 
 the best fit to the observed relation.  
This is, however, difficult in practise, because a change in $\log g$ can be compensated for 
by varying $\feddth$.

\begin{figure}
\begin{center}
\includegraphics[width= 0.73\columnwidth]{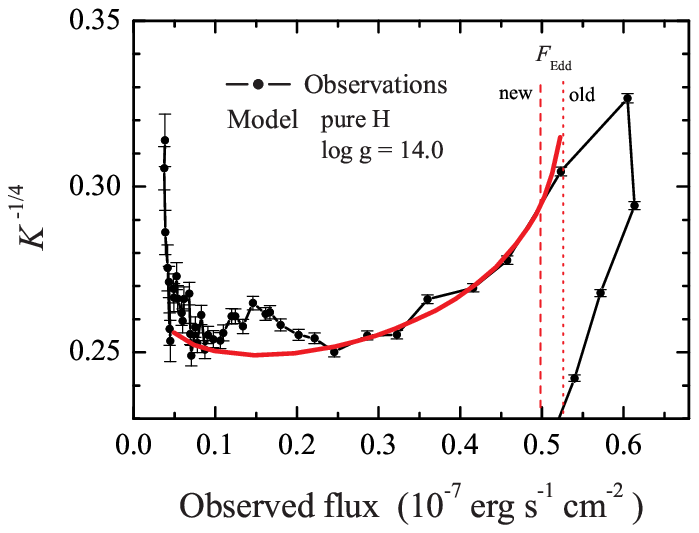}
\includegraphics[width= 0.73\columnwidth]{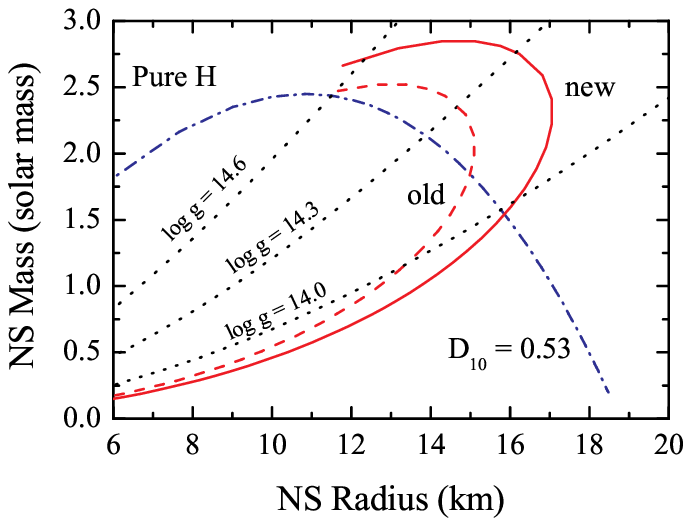}
\caption{\label{fig:13}
{\it Top panel:} The fit of the X-ray burst data for 4U\,1724--307 by the 
theoretical models for the NS atmosphere. 
The circles indicate the observed dependence of $K^{-1/4}$ versus $F$ for the long burst (S11) 
and the solid curve corresponds to the best-fit  theoretical model for a pure hydrogen atmosphere 
and $\log g=14.0$. 
Vertical dotted and dashed lines show the best-fit value for $\feddth$ for the old and new models. 
{\it Bottom panel:} Constraints on mass and radius of the NS in  4U\,1724--307. 
The new (solid) and the old (dashed) curves corresponding to $\teddinf$  are shown. 
The dash-dotted curve corresponds to the new best-fit parameter $A$ for the distance to the source of 5.3 kpc.  
  } 
\end{center} 
\end{figure}

The theoretical $\fc$--$l$ curve for $\log g = 14.0$ gives the best fit to the data 
(Fig.~\ref{fig:13}) for $A=0.168$ and  $\feddth = 4.93 \times 10^{-8}$ erg s$^{-1}$ cm$^{-2}$. 
The corresponding old, Kompaneets-based values 
are 0.170 and $5.25 \times 10^{-8}$, respectively. 
This small change in the best-fit parameters leads to a small decrease of $\teddinf$ 
from $1.64 \times 10^7$ K to $1.60 \times 10^7$ K and  a corresponding  increase in
 the NS radii (see Fig.~\ref{fig:13}, {\it bottom panel}).  
Interestingly,  the new $\teddinf$ curve  crosses with the curve $\log g$ = 14.0 at the NS mass 
close to 1.5$M_{\odot}$, and  
the curve $A=0.168$ also passes through the same crossing point  
when $D_{10}=0.53$ (i.e. 5.3 kpc, the distance to Terzan~2, \citealt{Ort:97}). 
The uncertainties in the obtained values arising owing to uncertainties in the data 
are very close to the differences between the old and the new values of the best-fit 
parameters (see Table 1 in S11).

A few attempts to find  NS mass and radius from the direct fitting of the observed X-ray burst spectrum by 
 NS atmosphere model spectra were performed \citep{MajMad05AcA,Miller:11,Kusm:11}.
They have proposed finding the best fit for a fixed $\log g$ by varying 
the relative luminosity $l$ (or effective temperature $T_{\rm eff}$) and the gravitational redshift $z$.  
The best-fit between all trial $\log g$ values then gives the desired $\log g$ and $z$. 
Thereafter,  the NS mass and radius can be found using Eqs.~(\ref{eq:g_def}) and (\ref{eq:redshift_def}). 
Unfortunately, the curves on the $M$--$R$ plane corresponding to the fixed values of $\log g$ and $z$ 
cross at very small angles (see Fig.~1 in SPW11). 
Therefore, the  uncertainties are expected to be large. 
More importantly, the model spectra are very close to the blackbody 
in the observed  {\it RXTE}/PCA energy band. 
This means that for an arbitrary product $\fc T_{\rm eff}$ (i.e. color temperature of the theoretical spectrum)  
we can find the redshift $z$ that makes the combination $\fc T_{\rm eff}/(1+z)$  
equal to the observed color temperature $T_{\rm bb}$, and in turn this method extremely unreliable.

\section{Summary}
\label{sec:summary}

We  have considered hot NS model atmospheres taking into account Compton scattering using the 
relativistic 
kinetic equation with the exact angle-dependent RF and  cross-section \citep{NP94,PS96},
accounting also for the induced scattering.  
We have developed a method to solve the obtained radiation transfer equation using a 
short characteristic method 
with the accelerated $\Lambda$-iteration. 
We have implemented this solution in our computer code for the model atmosphere calculations (SPW11). 

We have examined the properties of the  new model atmospheres.  
The main difference in comparison with the old models computed with  the Kompaneets 
operator concerns radiative acceleration. 
In the new approach, the Klein-Nishina reduction in the electron scattering cross-section  
leads to a decrease in the radiative acceleration  relative to that for  the Thomson  scattering cross-section.  
The grid of model atmospheres was extended to higher effective temperatures up to  luminosities formally  
exceeding the Eddington limit $\leddth$ computed for the Thomson cross-section. 
The role of the radiation pressure reduces in the deeper, hot, optically thick layers due to the same effect.   

We computed a new set of 484 hot NS model atmospheres. 
Following SPW11, we computed the models for six different chemical compositions 
(pure hydrogen, pure helium, solar hydrogen/helium mix with various abundances of heavy elements
 $Z$ = 1, 0.3, 0.1, and 0.01 $Z_{\odot}$) and three surface gravities ($\log g =14.0, 14.3$, and 14.6). 
The relative luminosities range from $l$=0.001 to 1.06--1.1 (depending on $\log g$). 
The new models with $l < 0.8$ are almost identical to the old models based on 
the Kompaneets operator approach. 
At higher $l$, the deviations are more significant due to a different treatment of the radiative acceleration. 
The difference between the new and old models is small if instead of $l$ one uses the relative 
radiative acceleration $g_{\rm rad}/g$ 
as a parameter. 

The spectra of the new and old models deviate by less than 5\% in the region around the spectral peak, 
and the deviation grows at higher energies. We have also tested various approximate and angle-averaged 
RFs for Compton scattering and found that they produce spectra that deviate from the exact 
solution typically by less than 2\%. The comparison of our spectra to the models computed using Madej's code 
\citep{madej:91,Madej.etal:04,Majczyna:05}  revealed dramatic differences in the spectral shape. 
We attribute these differences to their usage of the incorrect RF from \citet{Guilbert81}
and/or to the non-convergence of their radiation transfer calculations. 

The spectra of all our models were fitted by the diluted blackbody spectra in 
the {\it RXTE}/PCA energy band (3--20 keV) using 
the same five fitting procedures as in SPW11. 
The new $\fc$--$l$ relations are similar in shape to the old relations, 
but depend significantly on the surface gravity. 
However, the dependences of the color correction on $g_{\rm rad}/g$ are almost indistinguishable, with the 
new $\fc$ being larger by  approximately 1\%.  
The color corrections with the corresponding dilution factors, the theoretical
emergent spectral energy distributions, 
and specific intensities at various angles for all models are available at CDS.

In the present paper, we have also tested how the new  $\fc$--$l$ relations 
affect the determination of the NS masses and radii derived from the  spectral evolution observed during 
the cooling stages of PRE bursts. 
We found that the best-fit NS radius increases by about 10\% from the 
value obtained using the old model based on the Kompaneets equation (S11). 
We note that bursting NSs are rapidly rotating and accounting for the latitudinal variation in
gravity and the Doppler effect will affect the estimated NS radius.  This will be a subject of a separate publication.

\begin{acknowledgements} 
The work is supported by the German Research Foundation 
(DFG) grant SFB/Transregio 7 ÒGravitational Wave AstronomyÓ, 
Russian Foundation for Basic Research (grant 12-02-97006-r-povolzhe-a), 
the Jenny ja Antti Wihuri foundation, and  the Academy of  Finland (grant 127512). 
We also acknowledge the support of the International Space Science Institute (Bern, Switzerland), 
where part of this investigation was carried out.       
We are grateful to Jerzy Madej and Agata R{\'o}{\.z}a{\'n}ska for kindly providing us with 
the results of their atmosphere calculations. 
\end{acknowledgements}




\appendix

\section{Relativistic kinetic equation for Compton scattering and the RFs} 
\label{sec:rke}

For completeness let us rederive here the exact relativistic expressions for the Compton scattering RFs.  
For the detailed derivation, see \citet{NP93} and \citet{PV10}, where more general problems have been solved.  
In the first paper, the redistribution matrices  describing Compton scattering of 
polarized radiation in terms of Stokes parameters were derived, 
while in the second the RFs for anisotropic electron distribution have been obtained. 
We start from the relativistic kinetic equation (RKE) for photons that describes Compton scattering. 
 
\subsection{Radiative transfer equation}
\label{sec:rte}

A description of interactions between photons and electrons via Compton scattering accounting for 
the induced scattering and electron degeneracy 
can be provided by the explicitly covariant RKE for photons \citep{BY76,dGvLvW80,NP93,NP94}
\bea \label{eq:rke}
\fourx \cdot \unb \noccx(\vecx) &=&  \frac{\re^2}{2}  \frac{2}{\lambdac^3}
\int \frac{\rmd \vecp}{\gamma} \frac{\rmd \vecp_1}{\gamma_1}  \frac{\rmd \vecx_1}{x_1} 
\: F  \: \delta^4(\fourp_1 + \fourx_1 - \fourp - \fourx)   
\nonumber \\ 
&\times & 
\left\{ \noccx(\vecx_1) [1+\noccx(\vecx)]  \nocce(\vecp_1) [1-\nocce(\vecp)] \right. \\
&-& \left. \noccx(\vecx) [1+\noccx(\vecx_1)] \nocce(\vecp) [1-\nocce(\vecp_1)] ) \right\} , \nonumber
\eea
where $\unb=\{\partial/c \partial t, \vnabla\}$ is the four-gradient, 
$\re$ is the classical electron radius, $\lambdac=h/\me c$ is the Compton wavelength, 
$F$ is the Klein--Nishina reaction rate \citep{LLVol4}
\begin{equation} \label{eq:kn}
F  = \left( \frac{1}{\xi} - \frac{1}{\xi_1}\right)^2 + 2 \; \left( \frac{1}{\xi} - \frac{1}{\xi_1}\right)
+ \frac{\xi}{\xi_1} + \frac{\xi_1}{\xi} , 
\end{equation}
and 
\begin{equation} \label{eq:xixi1}
\xi = \fourp_1\cdot\fourx_1= \fourp\cdot\fourx, \qquad \xi_1 =  \fourp_1\cdot\fourx =  \fourp\cdot\fourx_1
\end{equation}
are the four-products of the corresponding momenta 
(second equalities in Equations (\ref{eq:xixi1}) arising  from the four-momentum conservation law represented 
by the delta-function in Eq. (\ref{eq:rke})). 
Here we define the dimensionless photon four-momentum as $\fourx =\{ x, \vecx \}= x \{ 1,\vomega\}$, 
where $\vomega$ 
is the unit vector in the photon propagation direction and $x \equiv h\nu/\me c^2$. 
The photon distribution  is described by either the occupation number $\noccx$ or  the specific intensity 
(per dimensionless 
energy interval) $I(\vecx)=x^3\noccx(\vecx)/C$, where the constant $C$ is given by Eq. (\ref{eq:const_int}).  
The dimensionless electron four-momentum is $\fourp  = \{ \gamma, \vecp\}= \{ \gamma, p\vOmega\} = 
\gamma \{ 1 , \beta\vOmega\}$, 
where $\vOmega$ is the unit vector along the electron momentum, 
$\gamma$ and $p=\sqrt{\gamma^2-1}$  are the electron Lorentz factor and its momentum in units of 
$\me c$, and $\beta$ is the velocity in units of $c$.
The electron distribution is described by the occupation number $\nocce$. 
For the isotropic electron distribution, we use  the electron distribution function  
$\fe(p)=2\nocce(\vecp)/\lambdac^3\Ne$, normalized to unity  
\be
4\pi \int_0^\infty \fe(p) \ p^2 \rmd p  = 1 . 
\ee

In the following, we consider a steady state 
and ignore electron degeneracy, because in the upper atmosphere layers, 
where the radiation spectrum is formed, electrons are non-degenerate. 
We define the RF as 
\be \label{eq:rf_gen}
R(\vecx_1 \rightarrow \vecx) = \! \frac{3}{16\pi}   
\int \!\! \frac{\rmd \vecp}{\gamma} \frac{\rmd \vecp_1}{\gamma_1}  \fe(p_1) F \delta^4(\fourp_1 + 
\fourx_1 - \fourp - \fourx)  .
 \ee
For the relativistic Maxwellian distribution of temperature $\Theta=k\Te /\me c^2$   
\citep{Jut11,synge57},
\be \label{eq:maxwell}
\fe(p) = \frac{1}{4\pi\ \Theta\ K_2(1/\Theta)} \exp(-\gamma/\Theta)
\ee
(where $K_2$ is the modified Bessel function),  
the RF  satisfies the symmetry property 
\be\label{eq:rf_symm}
R(\vecx\rightarrow \vecx_1) \ e^{-x/\Theta} = R(\vecx_1 \rightarrow \vecx) \ e^{-x_1/\Theta} , 
\ee
which follows from its definition in Eq. (\ref{eq:rf_gen}) and the energy conservation 
$\gamma_1=\gamma+x-x_1$, 
or from the detailed balance  condition (see eq. 8.2 in \citealt{Pom73}). 
Using this result it is easy to show that the Bose-Einstein distribution 
$\noccx(x)=1/(\exp\{[x-\mu]/\Theta\}-1)$
with any chemical potential is a solution of  the RKE (\ref{eq:rke}). 

In the absence of strong magnetic field, the medium is isotropic, therefore 
the RF depends only on the photon energies and the scattering angle (where $\eta$ is its cosine), 
i.e. we can write $R(\vecx_1\rightarrow \vecx) = R(x,x_1,\eta)$. 
The kinetic equation (\ref{eq:rke})  can then be recast in a standard form of the radiative transfer equation
\bea \label{eq:rte2}
\lefteqn{ 
\frac{\vomega \cdot \vnabla \noccx(\vecx) }{\sigmat \: \Ne}  \!= \!
- \noccx(\vecx) \frac{1}{x} \int_{0}^{\infty} \!\! x_1 \rmd x_1 \int \rmd ^2 \vomega_1 \: R(x_1,x,\eta)   [1+\noccx(\vecx_1)] 
 } \nonumber \\ 
& + &  [1+\noccx(\vecx)] \frac{1}{x}   \int_{0}^{\infty} \!\! x_1 \rmd x_1\!\! \int \!\! \rmd ^2 \vomega_1 \: R(x,x_1,\eta) \noccx(\vecx_1) .
\eea
For the plane-parallel atmosphere, this reduces to 
\bea \label{eq:rte_plane}
\lefteqn{ 
\mu \frac{\rmd \noccx(x,\mu)}{ \rmd \taut}   
=  \noccx(x,\mu) \frac{1}{x} \int_{0}^{\infty}  \!\!\!\! x_1 \rmd x_1 \!\!  \int_{-1}^{1} \!\!  
\rmd \mu_1 \: R(x_1,\mu_1;x,\mu)   [1+\noccx(x_1,\mu_1)]  }
\nonumber \\ 
& - & 
  [1+\noccx(x,\mu)] \frac{1}{x} \int_{0}^{\infty} \!\!\!\! x_1 \rmd x_1\!\!  \int_{-1}^{1} \!\! 
\rmd \mu_1 \: R(x,\mu;x_1,\mu_1)\noccx(x_1,\mu_1) ,
\eea
where $\rmd\taut=-\sigmat \Ne \rmd s= \kappae \ \rmd m$, 
$\mu$ and $\mu_1$ are the cosines of the angle relative to the normal, 
$\eta=\mu\mu_1+\sqrt{1-\mu^2}\sqrt{1-\mu_1^2}\cos\varphi$ and 
\be
R(x,\mu;x_1,\mu_1) = \int_0^{2\pi}  R(x,x_1,\eta) \ \rmd \varphi 
\ee
is the azimuth-integrated RF. 
Rewriting  Eq. (\ref{eq:rte_plane})  in terms of the intensity $I(x,\mu)$, we get the radiative transfer equation  (\ref{eq:rte}) 
that accounts for electron scattering with the scattering opacity and the source function given by 
Eqs. (\ref{eq:scatopac}) and (\ref{eq:source}), respectively.

\subsection{Redistribution functions}
\label{sec:rf}

The expression (\ref{eq:rf_gen}) for the RF can be simplified 
by taking the integral over $\vecp$ with the help of  the three-dimensional delta-function
and using the identity $\delta(\gamma_{1}+x_1-\gamma-x)=\gamma 
\delta \left( \fourx_1 \cdot \fourp_1- \fourx \cdot (\fourp_1+ \fourx_1) \right)$
\be \label{eq:rf_spe}
R(x,x_1,\eta) =  \frac{3}{16\pi}    \int  
\frac{\rmd \vecp}{\gamma} \fe(p) F \delta(\Gamma) ,
\ee
where we have dropped the subscript 1 from the electron quantities and 
\bea
\Gamma& =& \gamma(x_1-x) - p (x_1 \vomega_1 -x \vomega ) \cdot \vOmega - q , \\ 
q &=&\vecx\cdot\vecx_1 = xx_1(1-\eta) . 
\eea
To integrate over angles in Eq. (\ref{eq:rf_spe}), we 
follow the recipe proposed by \citet{AA81} (see also \citealt{PKB86,PV10}), 
choosing the polar axis along the direction of the transferred momentum
\be 
\vn \equiv \left( x_1 \vomega_1 -x \vomega\right) / Q ,
\ee
where 
\be \label{eq:Qcap}
Q^2=(x_1 \vomega_1 -x \vomega)^2= 
(x-x_1)^2+2q. 
\ee 
Thus, the integration variables become $\cos\alpha = \vn\cdot\vOmega$ and the azimuth $\Phi$. 
The RF (\ref{eq:rf_spe}) can then be written as
\be\label{eq:red_sim}
R(x,x_1,\eta)  \!= \! \frac{3}{16\ \pi}    \!\int\limits_{1}^{\infty} \!\! \, \fe (p)\ p\ \rmd \gamma 
\!\! \int \limits_{0}^{2\pi}  \rmd\Phi  
\int \limits_{-1}^{1} \!\! \, \delta(\Gamma)\  F \  \rmd\cos\alpha ,   	
\ee
where now
\be 
\Gamma= \gamma(x_1-x) - q -  p Q \cos\alpha .
\ee 
Integrating   over $\cos\alpha$ using  the delta-function, we get 
\be \label{eq:red_phi}
R(x,x_1,\eta)   =    \frac{3}{8}  \int_{ \gamma_{*}}^{\infty} \: \fe (p)\  R(x,x_1,\eta,\gamma)\ \rmd \gamma ,
\ee
where the integration over the electron distribution can been done numerically. 
Here we introduce the RF for monoenergetic electrons 
\be\label{eq:rfmono_def}
R(x,x_1,\eta,\gamma) = \frac{1}{Q} \frac{1}{2\pi} \int_{0}^{2\pi} \: F\ \rmd\Phi  . 
\ee
We need to substitute 
\be \label{eq:costheta}
\cos\alpha= \frac{\gamma(x_1-x) - q}{pQ} 
\ee
into the expression for $F$. The condition $|\cos\alpha|\le1$ provides the constraint 
\be\label{eq:gammastar}
\gamma \ge \gamma_* (x,x_1,\eta) = \left(x-x_1 +Q\sqrt{1+2/q} \right)/2 .
\ee
Integrating over azimuth $\Phi$ in Eq. (\ref{eq:rfmono_def}) gives the exact analytical 
expression for the RF that is  valid for any photon and electron energy \citep{AA81,NP94,PV10}, 
which we use in our calculations
\be\label{eq:r0f}
R(x,x_1,\eta,\gamma) \! = \!
\frac{2}{Q} + \frac{q^2-2q-2}{q^2} \left( \frac{1}{a_-}  \! - \!\frac{1}{a_+} \right)
+ \frac{1}{q^2} \left( \frac{d_-}{a_-^3}  \! + \! \frac{d_+}{a_+^3} \right), 	
\ee
where 
\bea
a_-^2 & =& (\gamma-x)^2 + \frac{1+\eta}{1-\eta}    , \quad 
a_+^2 = (\gamma+x_1)^2 + \frac{1+\eta}{1-\eta}  ,  \nonumber \\
d_{\pm} & = & \left( a_+^2 -  a_-^2 \pm Q^2\right)/2 . 
\eea
We note that the RF (\ref{eq:rfmono_def}) satisfies the detailed balance condition \citep{NP94}
\be\label{eq:rfmono_bal}
R(x_1,x,\eta,\gamma)  = R(x,x_1,\eta,\gamma+x-x_1).  
\ee

The RF (\ref{eq:rf_spe}) is related to the scattering kernel (8.13) of \citet{Pom73} as 
$R(x,x_1,\eta)= \sigma_s(x_1\rightarrow x,\eta)x_1/x$. 
The form given by Eq. (\ref{eq:red_sim}) is equivalent to eq. (A4) of \citet{BY76}. 
The derived RF  for monoenergetic electrons (\ref{eq:r0f}) 
is equivalent to eq.  (A5) of \citet{BY76} and eq. (14) of \citet{AA81}. 

A very simple approximate expression for the RF can be obtained by
assuming that the scattering in the electron rest frame proceeds in the Thomson regime (i.e. coherent) 
and is isotropic. This is equivalent to substituting $F$ by  $4/3$ in Eq. (\ref{eq:rfmono_def}). 
We then get \citep{an80,Pou94PhD,PS96} 
\be \label{eq:r0_app}
R(x,x_1,\eta,\gamma) \! = \! \frac{4}{3Q} . 
\ee
Integrating it over the Maxwellian distribution (\ref{eq:maxwell}) gives
\be \label{eq:r0tot_app}
R(x,x_1,\eta) = \frac{1}{8\pi Q} \frac{e^{-\gamma_*/\Theta}}{K_2(1/\Theta)} . 
\ee
This approximate RF is also used in the calculations. 
We note that this RF also satisfies the detailed balance condition (\ref{eq:rf_symm}).

\section{Method for solving  the radiation transfer equation}
\label{sec:method_rte}

The formal solution of the radiative transfer equation gives a relation 
between  the outward $I^+(x,\mu)$  and the inward $I^-(x,\mu)=I(x,-\mu)$ intensities at some depth point $i$ 
(on the optical depth grid $\tau^{\pm}_i, i=1,...,N$)  with the adjacent intensities 
\bea
    I^+_i(x,\mu) &=& I^+_{i+1}(x,\mu) \exp[-(\tau^+_{i+1}-\tau^+_i)/\mu] +
    \\ \nonumber
 &&   \int^{\tau^+_{i+1}}_{\tau^+_i}\,S^+(t,x,\mu)\exp[-(t-\tau^+_i)/\mu]\,dt/\mu,
    \\  
    I^-_i(x,\mu) &=& I^-_{i-1}(x,\mu) \exp[-(\tau^-_{i}-\tau^-_{i-1})/|\mu|] +
    \\ \nonumber
&&    \int^{\tau^-_{i}}_{\tau^-_{i-1}}\,S^-(t,x,\mu)\exp[-(\tau^-_i-t)/|\mu]\,dt/|\mu|.
    \eea
The integrals can be replaced by the sums using the parabolic approximation
\bea
    I^+_i(x,\mu) &=& I^+_{i+1}(x,\mu) \exp(-\Delta\tau^+_{i+1}) +
    \\ \nonumber
    &&\alpha^+_i S^+_{i-1}(x,\mu) +\beta^+_i S^+_i (x,\mu) + \gamma^+_i S^+_{i+1}(x,\mu),
    \\  
    I^-_i(x,\mu)& = &I^-_{i-1}(x,\mu) \exp(-\Delta\tau^-_{i}) +
    \\ \nonumber
 &&   \alpha^-_i S^-_{i-1}(x,\mu) +\beta^-_i S^-_i(x,\mu)  + \gamma^-_i S^-_{i+1}(x,\mu),
    \eea
where 
\bea
        \Delta\tau^{\pm}_i &=& (\tau^{\pm}_{i}-\tau^{\pm}_{i-1})/|\mu|,
    \\ \nonumber
      \alpha^-_i&=& e^-_{0,i} +
 \frac{e^-_{2,i}-(\Delta\tau^-_{i+1} +2\Delta\tau^-_{i})\,e^-_{1,i}}{\Delta\tau^-_{i}(\Delta\tau^-_{i+1}+\Delta\tau^-_{i})},
    \\ \nonumber
      \beta^-_i&=& ([\Delta\tau^-_{i} +\Delta\tau^-_{i+1}]\,e^-_{1,i}-e^-_{2,i})/(\Delta\tau^-_{i}\Delta\tau^-_{i+1}),
    \\ \nonumber
      \gamma^-_i&=& (e^-_{2,i}-\Delta\tau^-_{i}\,e^-_{1,i})/[\Delta\tau^-_{i+1}(\Delta\tau^-_{i+1}+\Delta\tau^-_{i})],
    \\ \nonumber
      \alpha^+_i&=& (e^+_{2,i+1}-\Delta\tau^+_{i+1}\,e^+_{1,i+1})/(\Delta\tau^+_{i}[\Delta\tau^+_{i+1}+\Delta\tau^+_{i}]),
    \\ \nonumber
      \beta^+_i&=& ([\Delta\tau^+_{i} +\Delta\tau^+_{i+1}]\,e^+_{1,i+1}-e^+_{2,i+1})/(\Delta\tau^+_{i}\Delta\tau^+_{i+1}),
    \\ \nonumber
      \gamma^+_i&=& e^+_{0,i+1} +\frac{e^+_{2,i+1}-(\Delta\tau^+_{i} +2\Delta\tau^+_{i+1})\,e^+_{1,i+1}}{\Delta\tau^+_{i+1}(\Delta\tau^+_{i+1}+\Delta\tau^+_{i})},
\eea        
and
\bea
      e^{\pm}_{0,i} &=& 1 - \exp(-\Delta\tau^{\pm}_{i}),
    \\ \nonumber
    e^{\pm}_{1,i} &=& \Delta\tau^{\pm}_{i} -  e^{\pm}_{0,i},
    \\ \nonumber
    e^{\pm}_{2,i}& =& (\Delta\tau^{\pm}_{i})^2 -  2e^{\pm}_{1,i}.
\eea
At the first depth point, the coefficients  are  
\be
 \alpha^+_1 = 0,  \quad \beta^+_1 =  e^{+}_{1,2}/\Delta\tau^+_{2}, \quad \gamma^+_1 = e^{+}_{0,2} - \beta^+_1,
\ee
and at the last point they are  
\be
\alpha^-_N = e^{-}_{0,N} - \beta^-_N, \quad   \beta^-_N =  e^{-}_{1,N}/\Delta\tau^-_{N}, \quad \gamma^-_N = 0.
\ee
We note that the inward and the outward opacities along the same ray are different (see Eq.~(\ref{eq:scatopac})).
 
 \begin{figure}
\begin{center}
\includegraphics[width= 0.73\columnwidth]{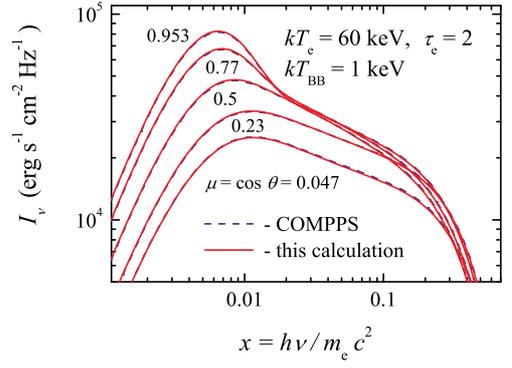}
\caption{\label{fig:b1}
 Comparison of the emergent spectra from a hot electron slab back-irradiated 
 by a blackbody computed using the {\sc compps} code and the code presented here. } 
\end{center} 
\end{figure}

The formal solution for a given  source function starts for the inward intensities from 
the outer boundary condition 
(the lack of incoming radiation at the surface)
 \be
     I^-_1(x,\mu) = \beta^-_1 S^-_1(x,\mu) + \gamma^-_1 S^-_2(x,\mu)
\ee
up to the last depth point $N$. 
The intensities at the innermost depth point are found using the inner boundary condition, which 
is taken from the diffusion approximation
\be
    I^+_N(x,\mu) = I^-_N(x,\mu) + 2 \frac{B_{x,N}-B_{x,N-1}}{\Delta\tau_N}.
\ee
The full solution is found iteratively using an accelerated $\Lambda$-iteration. 
At the first iteration,  the thermal part of the source function is taken.
For the subsequent iteration $n$, the intensities obtained from the previous iteration $n-1$ 
are used to compute the current source functions $S^{\pm,n}_i$.  
Iterations are continued until the relative change becomes smaller than the predetermined accuracy 
\be
    \max \left[\frac{J^{n}_i(x)}{J^{n-1}_i(x)}-1\right] < 10^{-4}, 
\ee
where $J_i(x)$ are  the mean intensities. 
This  solution  method of the radiation transfer equation was tested for a rather optically thin 
(Thomson optical depth $\tau_{\rm T} =2$)
and hot ($k\Te $ = 60 keV) electron slab back-illuminated by soft blackbody photons of
 $kT_{\rm BB}$ = 1 keV.
The solution for the emergent intensities obtained at five angles using our method were 
compared with the solution obtained with the Comptonization 
code  {\sc compps}  (\citealt{PS96}, see Fig.~\ref{fig:b1}).

 \begin{figure}
\begin{center}
\includegraphics[width= 0.73\columnwidth]{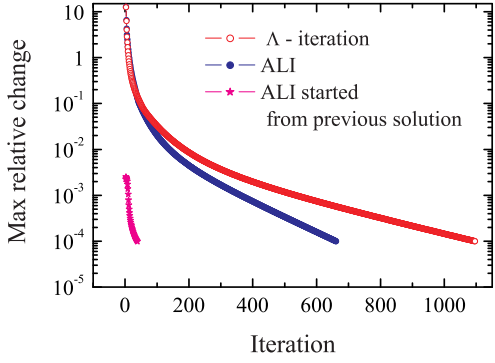}
\caption{\label{fig:b2}
The maximum relative change in the solution of the radiation transfer equation 
at the final temperature correction  computed by different versions of the accelerated $\Lambda$-iteration. 
} 
\end{center} 
\end{figure}

In the optically thick case ($\tau_{\rm T} \gg 1$), which is typical of NS atmospheres, 
the convergence of the solution can be accelerated using the following procedure. 
The difference between the formal solution obtained in the current iteration 
$ I^{\pm,n,\rm FS}$ and the solution at iteration $n-1$ is increased by some factor  
\be
       I^{\pm,n}_i(x,\mu) - I^{\pm,n-1}_i(x,\mu) = 
 \frac{ I^{\pm,n,{\rm FS}}_{i}(x,\mu)- I^{\pm,n-1}_i(x,\mu)}{1-\varepsilon^{\pm,n}_i(x,\mu)\,\Lambda^*_{i,i}(x,\mu)},
\ee       
 where
\be \label{u4a}
\varepsilon^{\pm,n}_i(x,\mu) = \frac{\sigma^{\pm,n}_i(x,\mu)}{\sigma^{\pm,n}_i(x,\mu)+k_i(x)},
\ee
and  $\Lambda^*_{i,i}(x,\mu)$ is the diagonal term of the approximate $\Lambda$--operator
\bea
    \Lambda^*_{i,i}(x,\mu) &=& \frac{1}{4}[\beta^+_i(x,\mu)+\beta^-_i(x,\mu)]\, \\
     &\times&  x^2\,\int_0^\infty
\frac{dx_{\rm 1}}{x^2_{\rm 1}} \int_{-1}^1 d\mu_{\rm 1} R(x, \mu;x_{\rm 1},\mu_1) . \nonumber 
\eea
The acceleration is not high (about 30 -- 40 \%) and the number of necessary 
iterations is still large (see Fig.~\ref{fig:b2}).  
However, in the process of the model atmosphere computation 
it is possible to use the source function from the previous temperature iteration as the starting approximation 
for the current source function (see details in Section\,\ref{s:methods}). 
In this case, the acceleration depends on the value of the temperature corrections $\Delta T_i$. 
At the  first few temperature iterations, when $\Delta T_i$ are large, the acceleration is insignificant, 
but at later iterations, when $\Delta T_i$ are relatively small, 
the accelerated $\Lambda$-iterations converge very quickly (Fig.~\ref{fig:b2}).

\section{Comparison with Madej's code}
\label{sec:madej}

\begin{figure*}
\begin{center}
\includegraphics[width= 0.73\columnwidth]{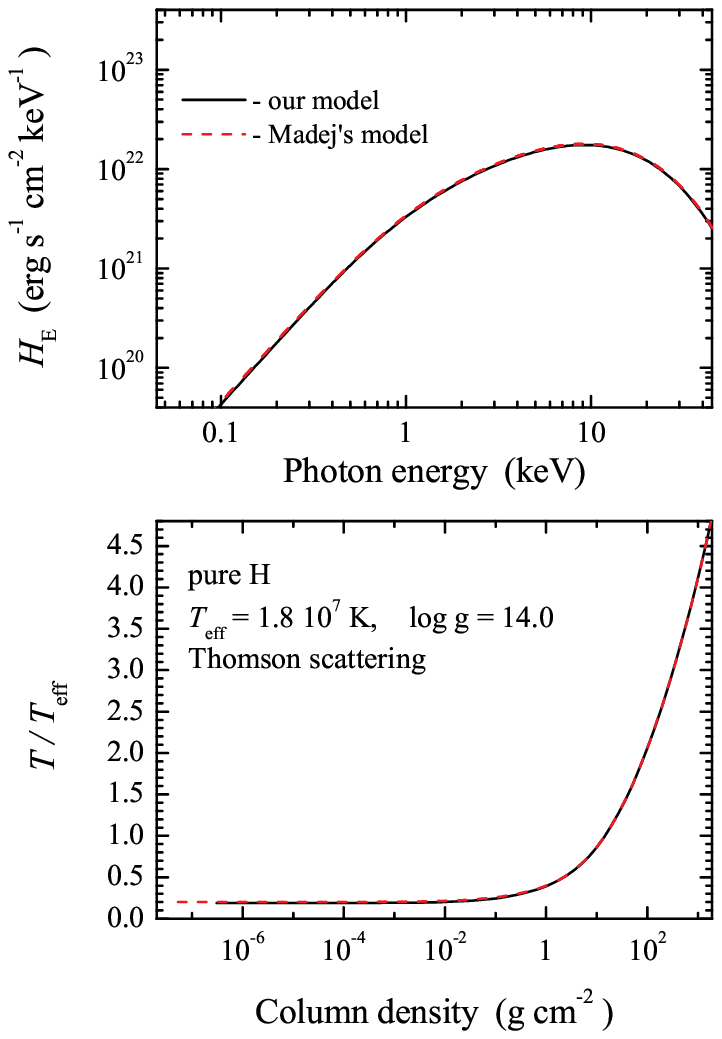}
\includegraphics[width= 0.73\columnwidth]{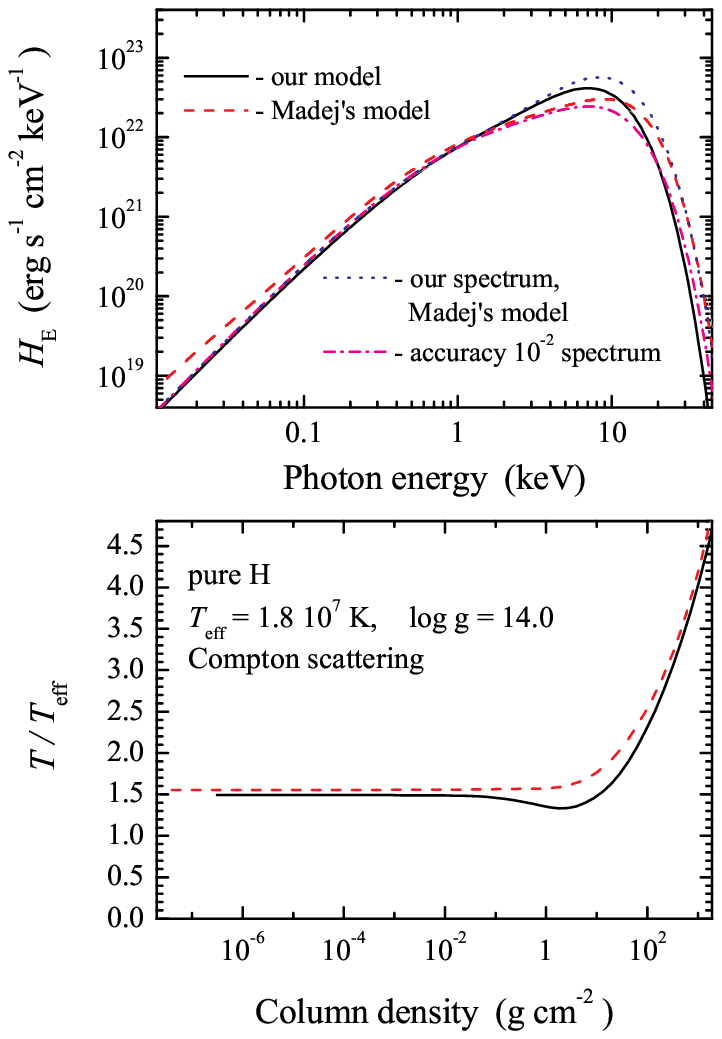}
\caption{\label{fig:madej}
{\it Left panels}: Comparison of emergent spectra  and temperature structures of the models 
computed by our code (solid curves) and  Madej's code 
(dashed curves),
when electron scattering is approximated by coherent Thomson scattering. 
{\it Right panels}: Same as left, but when the exact RF  was used to compute Compton scattering. 
The spectrum computed by us using the Madej's temperature structure is shown by the dotted curve 
and the spectrum computed with a relative accuracy of $10^{-2}$ is presented by  dash-dots.
} 
\end{center} 
\end{figure*}

The only other attempt to compute NS atmospheres using an 
integral approach to Compton scattering going beyond the 
Kompaneets approximation was that of  J. Madej and collaborators \citep{madej:91,Madej.etal:04,Majczyna:05}.
They used an angle-averaged RF for Compton scattering derived by \cite{Guilbert81}. 
It is important to compare our results with those obtained by Madej's code for the same input parameters.  
We  selected our fiducial model as a testbed and computed two models. 
In the first, we approximated electron scattering by coherent Thomson scattering. In the second, 
the exact fully relativistic RF for Compton scattering was used. 
Calculations for identical parameters 
were performed with Madej's code (J. Madej and A. R{\'o}{\.z}a{\'n}ska, private communication).  
A comparison between the results  is shown in Fig.\,\ref{fig:madej}. 

We see that the Thomson scattering models are very close to each other, in terms of
both the spectra and the temperature structures  (Fig.~\ref{fig:madej}, {\it left panels}). 
However, the models  with Compton scattering differ substantially (Fig.~\ref{fig:madej}, {\it right panels}). 
Temperatures in the upper layers with column densities less than $10^3$ g cm$^{-2}$  
are lower in our model by up to 5--15\%, 
and  our spectrum (solid curve) is much more peaked and softer than that computed by Madej's code
 (dashed curve). 
We also note that our spectrum is very close to the diluted blackbody spectra, which 
cannot be said about Madej's  spectrum. 

We  can suggest two hypotheses to explain the discrepancies. 
First, there is a difference in the  RF used by the two codes. 
We employed  the exact relativistic RF (see Appendix \ref{sec:rf}), which 
had been extensively studied  and tested  against Monte-Carlo simulations \citep{SPS95,ZPJ00}.
Madej and collaborators \citep{madej:91,Madej.etal:04,Majczyna:05} used 
the (angle-averaged) RF derived by \citet{Guilbert81} (see  his Eqs.  (8) and (10)), which 
differs from  the correct expression (\ref{eq:rfmono_def}) by an additional factor 
$[1-(\beta\cdot\vn)(\vn\cdot \vomega_1)]/(1-\beta\cdot \vomega_1)$, 
which appeared from an error in the Jacobian. 
Guilbert's RF also does not satisfy the detailed balance condition (\ref{eq:rfmono_bal})
and therefore the RF integrated over the electron distribution does not satisfy the condition 
(\ref{eq:rf_symm}) either. 
Because the energy transfer by Compton scattering scales as $\beta^2$, 
the error on the order of $\beta$ obviously invalidates all the results obtained with this RF. 

Second, some of the discrepancies could be connected to a difference in the 
computation of the radiation transfer 
and the atmosphere modeling. We illustrate this  in Fig.~\ref{fig:madej} ({\it top right panel}). 
Using our radiation transfer code, we computed the spectrum using the temperature structure from 
Madej's calculations.  When calculations proceeded until the accuracy of 10$^{-4}$ was reached (dotted curve), 
the spectrum was much above our benchmark spectrum, i.e. it had a much higher effective 
(and color) temperature. 
If calculations were stopped at a  lower accuracy of 10$^{-2}$ (dash-dotted curve),  the spectrum  
was found to be  similar to Madej's spectrum. This suggests that the radiation transfer iterations 
with Madej's code did not sufficiently converge.

Madej et al. use the partial linearization method described in \citet[][p. 179]{Mihalas:78} and modified
 to include Compton scattering. 
Their radiation transfer equation is rewritten to include the linearized part of the radiative equilibrium equation. 
The solution of this generalized radiation transfer equation satisfies simultaneously 
 the radiative equilibrium to the first order.
Using the computed radiation field,
 the  temperature correction  for the current temperature structure of the atmosphere is found. 
This procedure is iterated until some convergence criterion is satisfied, for example, 
the emergent bolometric flux is 
accurate to within 0.1 \%.  
In this method, the accuracy of the current solution of the radiation transfer equation is on
 the order $\Delta T / T$, 
which is about $10^{-2}$--$10^{-3}$ at the last iteration.
We have shown above that this internal accuracy is insufficient to obtain the exact solution 
of the radiation transfer 
equation when Compton scattering is taken into account.  
We note that using this method is not possible 
to solve the radiation transfer equation for a  given atmosphere model  or, for example, 
for a homogeneous isothermal slab. 
This means that it cannot be checked independently of the atmosphere modeling.

\section{Atmosphere model spectra and color-corrections}
\label{app:D}

Table D.1 gives the color-correction and dilution factors from 
the blackbody fits to 484 atmosphere model spectra (fluxes), 
which in their turn are given in Table D.2. 
Table D.3 contains the emergent specific intensities at three angles. 
Tables D.1, D.2 and D3 are only available in electronic form at the CDS via anonymous ftp to cdsarc.u-strasbg.fr (130.79.128.5) or via http://cdsweb.u-strasbg.fr/cgi-bin/qcat?J/A+A/.

\end{document}